\documentclass{aa}  
\usepackage{natbib}
\bibpunct{(}{)}{;}{a}{}{,} 
\usepackage{graphicx}
\usepackage{txfonts}
\usepackage{verbatim}
\begin{document}
   \title{The interstellar cosmic-ray electron spectrum  from synchrotron radiation and  direct measurements.}

   \subtitle{}

   \author{A. W. Strong
          \inst{1}
          \and
          E. Orlando\inst{2,1}
          \and
          T. R. Jaffe\inst{3,4}
          }

   \offprints{A. W. Strong}

\institute
{
Max-Planck-Institut f\"ur extraterrestrische Physik,
Postfach 1312, D-85741 Garching, Germany \\
\email{aws@mpe.mpg.de}
\and
W.W. Hansen Experimental Physics Laboratory, Kavli Institute for Particle
Astrophysics and Cosmology, Stanford University, Stanford, CA 94305, USA \\
\email{eorlando@stanford.edu}
\and
Universit\'e de Toulouse; UPS-OMP; IRAP;  Toulouse,  France
\and
 CNRS; IRAP; 9 Av. colonel Roche, BP 44346, F-31028 Toulouse Cedex 4, France
 }

   \date{Received 4 March 2011 / Accepted 17 August 2011 }

 
  \abstract
  {The relation between Galactic cosmic-ray electrons, magnetic fields and synchrotron radiation.}
   {  We exploit synchrotron radiation to constrain the low-energy interstellar electron spectrum, using various radio surveys and connecting with electron data from Fermi-LAT and other experiments.  }
{  The GALPROP programme for cosmic-ray propagation, gamma-ray and synchrotron radiation is used. Secondary electrons and positrons are included. Propagation models based on cosmic-ray and gamma-ray data are tested against synchrotron
data from 22 MHz to 94 GHz. }
{The synchrotron data  confirm the need for a low-energy break in the cosmic-ray electron injection spectrum. The  interstellar spectrum below a few GeV has to be lower than standard models predict, and this suggests less solar modulation than usually assumed. Reacceleration models are more difficult to reconcile with the synchrotron constraints.  We show that secondary leptons are important for the interpretation of synchrotron emission. We also consider a cosmic-ray propagation origin for the low-energy break.  }
  { Exploiting the complementary information on  cosmic rays and synchrotron gives unique and essential constraints on  electrons, and has implications for gamma rays. This connection is especially relevant now in view of the ongoing PLANCK and Fermi missions.}

   \keywords{cosmic rays Ð  Ð synchrotron radiation Ð magnetic fields Ð gamma rays }

\titlerunning{Cosmic-ray electron spectrum from synchrotron}
\maketitle
\def\Breg{B_{reg}}
\def\Bran{B_{ran}}
\def\Bperp{B_{perp}}
\def\Btot{B_{tot}}
\def\B{{\bf B\ }} 
\def\microG{$\mu G$\ }
\def\Rscale{\Delta R}
\def\zscale{\Delta z}


\section{Introduction}

  Direct measurements of cosmic-ray (CR) electrons extend from TeV down to 1 GeV (and lower energies from spacecraft like Ulysses and Voyager), but solar modulation complicates their intepretation at  energies below about 10 GeV.
 Fermi-LAT \citep{2009PhRvL.102r1101A,2010PhRvD..82i2004A} 
 has made the currently most precise electron measurements in the range 7 GeV - 1 TeV; here modulation is of less importance although still significant in the lower part of this range. Synchrotron radiation (from tens of  MHz to tens of GHz)  probes  interstellar electrons from 0.5 to 20 GeV for the typical Galactic magnetic field (hereafter B-field) of a few $\mu$G, and hence can be used in conjunction with direct measurements to construct the full spectrum from GeV to TeV. At low energies this will finally allow an independent estimate of solar modulation for testing heliopheric propagation models.
In contrast such a probe of the low-energy interstellar spectrum is not available for CR nuclei.

An extensive review of CR propagation including electrons can be found in \cite{2007ARNPS..57..285S}; 
 a recent global viewpoint for the Milky Way is given in \citet{2041-8205-722-1-L58}, and a propagation parameter study in \citet{2011ApJ...729..106T}. 
Further discussion of the electron spectrum measured by Fermi-LAT can be found in \cite{2010A&A...524A..51D,2011NIMPA.630...48G,2011APh....34..528D}. Its  relation to the high-energy positron excess discoved by the PAMELA  experiment  \citep{2009Natur.458..607A} is also discussed in these papers. Our preliminary study of the synchrotron spectrum using GALPROP was given in  \citet{2009arXiv0907.0553O}.

The main objective of the present paper is to constrain the interstellar electron spectrum using a combination of the latest electron spectrum measurements and synchrotron radiation, leaving the question of its origin via injection and propagation as a side-issue. Hence while we use the CR propagation code GALPROP  to generate interstellar  spectra for various propagation scenarios, the latter are not our  main focus. Nevertheless it is shown that some current models are actually excluded on the basis of the synchrotron data. The relation between synchrotron and electrons is complicated by the presence of secondary electrons and positrons, and hence these are included in our model and their effect is addressed.

In previous analyses we were able to constrain the total B-field $\Btot$ on the basis of synchrotron data
and the distribution of cosmic-ray electrons derived from gamma rays \citep{2004ApJ...613..962S};
  we obtained  a local value $\Btot$ = 6\microG and a scale length of 8~kpc in R and 1~kpc in z.
The value of $\Btot$  has frequently been quoted in the literature as an independent measurement
of the total B-field.
Since then the CR source distribution has been modified to better reflect the
distribution of SNR as traced by pulsars {\citep{2004A&A...422L..47S}, and this influences the
derived
radial variation of $\Btot$; the larger radial variation of the source function implies a smaller variation of  $\Btot$.
Also the new measurement of the electron spectrum by Fermi-LAT  \citep{2009PhRvL.102r1101A,2010PhRvD..82i2004A} and the elimination by Fermi-LAT of the EGRET gamma-ray `GeV excess'\citep{2009PhRvL.103y1101A}\footnote{our earlier work was based on an electron spectrum adjusted to fit this excess, and which was a factor 4 higher than locally measured,  while now we use the measured electron spectrum.}
  lead to an upward revision of the value of  $\Btot$. We concentrate on the spectral aspects of synchrotron emission in this paper.

The magnitude of the B-field is a free parameter in this analysis, since while the regular component can be determined from rotation measures of pulsars and extragalactic sources, this is only a fraction of the total field.  Our approach is to use the models for regular component derived from RMs, combining these with a random field to be determined, and the latter is one of the results of our analysis.

 Extensions to lower frequencies where absorption is important have been made, see for example 
 \cite{1978JPhG....4.1793S} 
 and
 \cite{2008JGRA..11311106W}. 


 In a paper complementary to this one, we also address the relation between cosmic-ray electrons, synchrotron and B-fields
\footnote{
\citet{2011arXiv1105.5885J}. 
 Both papers use the GALPROP code to model CR propagation.
 The main difference is that they consider emission in the Galactic plane from 408 MHz to 23 GHz, while we use data out of the plane down to 22 MHz.  They address the spatial variations induced by the  B-field, while we concentrate only on the spectral aspects. We explicitly compare the contributions from secondary leptons, while they consider only the total. They use polarization (with Faraday rotation) to separate the regular and random B-field components. They also use rotation measures to constrain the regular B-field.  Despite the differences in approach,  the conclusions -  on issues  common to the papers -  are consistent.
A recent paper describing a similar  approach to ours  \citep{2011arXiv1106.4821B} 
 has been brought to our attention.
}.


\section{GALPROP development}

A description of  the GALPROP\footnote{The code is publicly available at http://galprop.stanford.edu.} model can be found in \cite{2007ARNPS..57..285S} and references therein; in particular see \cite{1998ApJ...509..212S}, \cite{2004ApJ...613..962S} and \cite{2008ApJ...682..400P}, and the GALPROP Explanatory Supplement available from the GALPROP website.

A shortcoming of GALPROP up to now has been the simplified B-field modelling, using only
a random component and a simple 2D exponential dependence.
With the availability of excellent radio continuum surveys from tens of MHz to tens
of GHz, including the WMAP satellite data, more sophistication is desirable;
 we have therefore introduced full 3D models for both
regular and random B-field . 
For the present spectral study only the total field is required (derived from the 3D model), and a 2D propagation scheme is sufficient.


\subsection{Synchrotron emissivity calculations}

From the spectrum of particles (here electrons or positrons) computed by GALPROP at
all points on the 3D grid, 
 we integrate  over particle energy  to get the synchrotron
emissivity for the regular and random fields.

The emissivity as seen by an observer at the solar position is computed as a function of (x, y, z, $\nu$). The spectrum and distribution of the emissivity thus depends on the form of  the regular and random components of the magnetic field, and the spectrum and distribution of CR leptons.

\subsubsection{Regular field}

The synchrotron emissivity (in erg s$^{-1}$  Hz$^{-1}$ ) of an isotropic distribution of monoenergetic
relativistic particles in a uniform magnetic field has
polarized components parallel and perpendicular to the projection
of the field on the line-of-sight to the observer \citep{2010hea..book.....L}:
\begin{equation}
\epsilon_{par}(\nu) = \frac{\sqrt{3}}{2}~{e^{3}\over mc^{2}}B_{perp}~[F(x) - G(x)]
\end{equation}
\begin{equation}
\epsilon_{perp}(\nu)=  \frac{\sqrt{3}}{2}~{e^{3}\over mc^{2}}B_{perp}~[F(x) + G(x)]
\end{equation}
where x = $\nu/\nu_{c}$, with $\nu_{c}$= ${3\over4\pi}~{e\over mc}B_{perp}\gamma^{2}$ and with $\gamma$ the electron Lorentz factor, 
$B_{perp}= B(x,y,z)~ sin\alpha$, with $\alpha$ the angle between the magnetic field and the line-of-sight.
 The functions F(x) and G(x) are defined in terms of Bessel functions \citep{2010hea..book.....L}, with:
\begin{eqnarray}
F(x)=  x~\int_{x}^{\inf}~K_{5/3}(x') ~dx'\nonumber\\ 
G(x)=  x~K_{2/3}(x) 
\end{eqnarray}
where $K_{5/3}(x)$ and $K_{2/3}(x)$ are the modified Bessel functions of order 5/3 and 2/3.
They are conveniently provided as C library functions in 
the GNU Scientific Library\footnote{http://www.gnu.org/software/gsl}.
 The resulting synchrotron spectrum has a broad maximum centred roughly at the frequency $\nu_{c}$ and the maximum has a value $\nu_{max}$=0.29~$\nu_{c}$ \citep{2010hea..book.....L}.

The polarization formulation will be of use for our future work. Here we are only interested in the total intensity given by the sum of the two components described above:
\begin{equation}
\epsilon(\nu, \gamma)=  \sqrt{3}~{e^{3}\over mc^{2}}B_{perp}~F(x) 
\end{equation}

\subsubsection{Random field}

For a randomly oriented field the emissivity is isotropic and obtained by
integrating the regular field expressions over all solid angles.
The result is given by \citep{1988ApJ...334L...5G}
\begin{equation}
\epsilon_{rand}(\nu)= C\ x^2 [K_{4/3}K_{1/3} -{3\over5} x (K_{4/3}K_{4/3} -
K_{1/3}K_{1/3})]
\end{equation}
$x=\nu/\nu_c$,  $\nu_c={3\over2\pi} {e\over mc}\Bran\gamma^2$,
$C={2\sqrt3} {e^3\over mc^2} \Bran$ erg s$^{-1}$ Hz$^{-1}$,
and the Bessel functions $ K_{4/3}, K_{1/3}$ are again computed using the GNU
Scientific Library.

Our implementation has been checked by integrating the regular
field expression over solid angle, giving exact agreement
with this  formula.

\subsection{Synchrotron intensity}

With GALPROP calculation of emissivity on the grid, we  integrate
over the line-of-sight to get the synchrotron intensity
for the regular and random fields. 
The synchrotron intensity at frequency $\nu$ is then given by 
\begin{equation}
I(\nu)=\int \epsilon(\nu) ~ds
\end{equation}
The observed brightness temperature\footnote{ In this paper we use the temperature spectral index, as is commonly used in radioastronomy (e.g. literature in Appendix A), which is equal to to the intensity spectral index plus 2.}
 of the radiation seen in a given direction is 
\begin{equation}
T(\nu)\propto \frac{c^{2} I(\nu)}{2~\nu^{2}}
\end{equation}

The resulting synchrotron skymaps for a user-defined grid of frequencies are output
by GALPROP  in Galactic coordinates either as CAR (Carr\'e projection) or in HEALPix \citep{2005ApJ...622..759G} (the preferred format).
The emissivity as seen by an observer at the solar position is also output as a
function of $(R,z,\nu)$ (spatial 2D) or $(x,y,z,\nu)$ (spatial 3D).



\subsection{Galactic magnetic field models}

In the same way as for the other Galactic constituents (gas, ISRF, cosmic rays), the magnetic field
is defined on a grid, in 2D or 3D. Only the 3D case is relevant for the full B-field
model, although the 2D case is retained for compatibility.

In \citet{2000ApJ...537..763S} only the random component of the magnetic field 
was present and was implemented in 2D with an exponential law for the  component.

Since that work, many 3D models of the Galactic magnetic field have been implemented in GALPROP in order to calculate the synchrotron emission from the Galaxy. 
The regular B-field used in the present work is the  model RING-ASS of \citet{2008A&A...477..573S} 
 for the disk, based on rotation measures of extragalactic radio sources. This has typically $\Breg = 2\mu$G. A toroidal halo field  is also included as prescribed in
 \citet{2010RAA....10.1287S},  
having a typical  value of 2$\mu$G.
 We include the regular field in order to make our model compatible with current information, but in fact since it is much less than the random field, this is not critical to our study.

 From Section 2, $\nu_c= (E/\ 1\ GeV)^2\ (B_{ran}/7.5\mu G)\times$ 240~MHz, so that the full range of synchrotron frequencies used,  22 MHz to 94 MHz, traces electron energies from roughly 0.5 to 20 GeV for our adopted B-field.

At 408 MHz and above, secondary leptons (above a few GeV) become less important for synchrotron and the relevant leptons are measured directly without much solar modulation, while at higher frequencies ($>$1~GHz) free-free emission can start to enter, so this is the best frequency  to determine the random B-field. Our model for $\Bran$   is therefore based on fits to the data at 408 MHz.
 The random field is modelled as a double exponential in (R,z), the free parameters being the two scale lengths (30~kpc in R and 4~kpc in z) and the local B-field: 
$\Bran = 7.5\mu$G \footnote{We note that  \citet{2008A&A...477..573S}  use an electron spectrum a factor 3 higher than the Fermi-LAT measurements, and hence obtain a random field   lower than ours (3$\mu$G).}.
This model reproduces  the longitude and latitude distribution of synchroton at 408 MHz sufficiently well for our purpose.
Note that since we are mainly concerned with spectral shape in this paper, the
absolute value of the B-field and its spatial distribution are not critical, and affect  mainly the relation of electron  energy to synchrotron frequency.


\section{Cosmic-ray model}

We use the GALPROP models described in \citet{2041-8205-722-1-L58}, to which we refer for details.
These models have been adjusted so that the propagation parameters are consistent with CR nuclei secondary-to-primary ratios.
Only the electron injection spectrum and the B-field are varied with respect to these models; these do not affect the validity of the  propagation parameters.

Electrons and positrons lose energy by synchrotron radiation, and this is included
in GALPROP self-consistently using the $\Btot$ of the adopted model.
Energy loss by inverse Compton scattering, ionization, Coulomb  and bremsstrahlung are also included,
although the latter three processes are of minor importance at the electron energies of interest here (see \cite{1998ApJ...509..212S} for a plot of the loss processes).

We concentrate on the spectrum of electrons as determined by synchrotron data, combining these with direct measurements.
Electron spectra are modelled to be consistent with Fermi-LAT data in the range 7 GeV to 1 TeV, including an estimate of solar modulation which still has some effect at these energies.
Below 7 GeV the primary electron injection spectrum is  modelled with 2 breaks. The minimal constraint is that the propagated spectrum must not be below the directly-measured spectrum  (i.e. at least some solar modulation is present).
 Secondary electrons and positrons (from pp, p-He, He-p and He-He interactions via pion-decay)
 are included using the standard GALPROP treatment with locally-measured proton and Helium CR spectra as a normalization.
Note that since Fermi-LAT measures electrons plus positrons, our primary electrons source effectively includes the high-energy positrons primary component measured by PAMELA \citep{2009Natur.458..607A}, so the latter is not explicitly modelled here (see e.g. \citet{2011NIMPA.630...48G,2011APh....34..528D} for such  models).

 As usual all the GALPROP parameter files for the models will be available as Supplementary Material. The plots will be made available in numerical form on request.


\section{Radio data}

\subsection{Review of information on spectral indices}

A review of  radio continuum surveys and their calibration is given in  \citet{2009IAUS..259..603R}.
We are interested in both spectral and spatial properties of the synchrotron sky, but not  in the fine angular details.
For the spatial distribution the most useful is the full-sky 408 MHz survey \citep{1982A&AS...47....1H}, 
 since it has full sky coverage, a well-established calibration and zero level,
and is a standard in this context. It has also the advantage that contributions from non-synchrotron components (eg free-free emission) are rather small.
For spectral information we have assembled a set of surveys which are described in Section 4.2.

There are also many detailed observational studies aimed at measuring accurate sky temperatures and spectral indices at particular wavelengths, for particular sky regions. These are very valuable in providing absolute values, especially at high Galactic latitudes. They also address the question of the contribution from extragalactic radio sources.
We list some representative results from the literature on spectral indices in Appendix A.
In summary, there is a wealth of measurements showing that the spectral index of the synchrotron emission increases steadily from about 2.5 to 3.0 over the frequency range from tens of MHz to tens of GHz. While there is considerable scatter in the actual values, they still provide an essential observational constraint on  any model for the CR electron + positron spectrum. Here we use a representative sample to compare with our models, but the comparison is just indicative of the general trend since the experiments cover many different sky areas.


\subsection{Radio surveys used}

 We use surveys at frequencies from 22 MHz to 23 GHz (and up to 94 GHz for WMAP) to compare directly the synchrotron spectrum with the models.
The surveys used are summarized below. Some were obtained directly from their authors, others from the Bonn\footnote{http://www.mpifr-bonn.mpg.de/survey.html}  and LAMBDA\footnote{http://lambda.gsfc.nasa.gov} websites.
The combined zero level and extragalactic/CMB  corrections were taken from the literature as stated below.

{\bf 22~MHz}: DRAO Northern hemisphere survey: \citet{1999A&AS..137....7R}. 
{\bf 45~MHz}:  North: \citet{1999A&AS..140..145M}
, South: \citet{1997A&AS..124..315A}
, combined all sky: \citet{2011A&A...525A.138G}. 
This is complete apart from two regions of $10^o$ and $20^o$ radius out of the plane. 
 An offset of 550K was subtracted  \citep{2011A&A...525A.138G}.
{\bf 150~MHz}: Parkes-Jodrell Bank all sky survey: \citet{1970AuJPA..16....1L}. 
{\bf 408~MHz}: Bonn-Jodrell Bank-Parkes all sky survey: \citet{1982A&AS...47....1H}. 
An offset of 3.7K was subtracted \citep{1988A&AS...74....7R}\footnote{\citet{2004mim..proc...63R} give 2.7K, \citet{2011A&A...525A.138G}   give 1.6K,   but the difference is not significant here.}.
{\bf 1420~MHz}: Stockert-Villa Eliza all sky survey. North: \citet{1982A&AS...48..219R,1986A&AS...63..205R},  
 South: \citet{2001A&A...376..861R}. 
An offset~of 2.8K was subtracted \citep{2004mim..proc...63R}.
{\bf 2326 MHz}: Rhodes southern hemisphere survey:  \citet{1998MNRAS.297..977J}. 
This was not used here due to restricted coverage, but will be used when the analysis is extended to lower Galactic latitudes in future.
{\bf 23~-~94~GHz}: For  WMAP   we used the spectral-index maps generated on the basis of WMAP polarized data by \citet{2008A&A...490.1093M} , which we used to scale the 408 MHz map to the WMAP frequencies\footnote{ \citet{2008A&A...490.1093M} used the 408 MHz maps uncorrected for zero offset (Miville-Desch\^enes, private communication), so we use the same maps for consistency when applying their index maps.};
 in this way the synchrotron radiation is extracted essentially uncontaminated by thermal and spinning dust\footnote{Their re-analysis of the WMAP synchrotron data   including spinning dust correction  resulted in a lower intensity than previous analysis. Spinning dust emission is produced by  small grains rotating and produces unpolarized radio emission. They  analyzed the combination of the WMAP polarization and intensity data finding strong evidence for the presence of unpolarized
spinning dust emission in the 20-60 GHz range. They performed an analysis of the WMAP synchrotron emission at 23 GHz where the signal to noise ratio is the highest  and the polarised emission is only synchrotron. Their estimates of the intensity at this frequency are based on extrapolation of the Haslam 408 MHz data with a spatially varying spectral index constrained by the WMAP 23 GHz polarization data. Hence, supposing that the synchrotron spectral index does not vary with frequency over
the WMAP range, they found an anomalous emission
with a spectrum from 23 to 61 GHz in accordance with the
models of spinning dust.}.

The surveys used have full-sky  (or almost full-sky) coverage apart from 22 MHz, but  here the coverage is still almost complete in the sky region used.  The coverage of these surveys can be seen graphically in \citet{2008MNRAS.388..247D}\footnote{They give an extensive list of  radio surveys  including older ones. They also provide some of them in HEALPix and use a subset for a principal components analysis with the aim of predicting the radio sky at any frequency. However issues of calibration and zero-level are not addressed by them so we do not use their results here.}.  

In order to avoid absorption effects at low frequency, and free-free emission at higher frequencies, and to avoid effects of zero-level corrections and local emission,  this analysis is restricted to regions out of the Galactic plane but avoiding the polar regions, specifically $10^o<|b|<45^o$. The thermal contribution even at 1420 MHz is then small \citep{1988A&A...196..211R,1989MNRAS.237..381B},  
a recent estimate being 15\% \citep{2003MNRAS.341..369D}, 
and less at lower frequencies (while at WMAP frequencies our  maps separate out the  non-thermal component as explained above).
 We also avoid the North Polar Spur by avoiding the region  $340^o<l<40^o$, although this hardly affects the result over such large sky regions, as we have verified.

 It is worth noting that lower frequency data are available (e.g. down to 1.3 MHz from the RAE2 satellite, \cite{1978ApJ...221..114N}), discussed in \cite{1978JPhG....4.1793S}, but these are strongly affected by absorption so are not used here.

For the spectral comparisons, the surveys and models predictions were converted to HEALPix \citep{2005ApJ...622..759G} and averaged over the stated sky region. Since HEALPix has equal solid angle pixels the correct averaging is ensured.

\section{Electron and positron data}

The CR electron and positron data used are as follows:
 AMS01      \citep{2002PhR...366..331A},
 CAPRICE94  \citep{2000ApJ...532..653B},
 HEAT       \citep{2001ApJ...559..296D},
 SANRIKU    \citep{1999ICRC....3...61K},
 BETS       \citep{2001ApJ...559..973T},
 PPT-BETS   \citep{2008AdSpR..42.1670Y},
 ATIC-1-2   \citep{2008Natur.456..362C},
 H.E.S.S.   \citep{2008PhRvL.101z1104A,2009A&A...508..561A}, 
 Fermi-LAT  \citep{2009PhRvL.102r1101A,2010PhRvD..82i2004A}, 
 PAMELA     \citep{2011PhRvL.106t1101A}.
 The data are taken from the CR database  described in \citet{2009arXiv0907.0565S}
\footnote{available from http://www.mpe.mpg.de/$\sim$aws/propagate.html.}.

Fermi-LAT is taken as definitive above 7 GeV, while the extension to lower energies is based on AMS01, CAPRICE94 and HEAT.
The low-energy data were taken around solar minimum with corresponding modulation levels.
For  HEAT  the modulation levels quoted were 755 MV for 1994 and 670 MV for 1995 \citep{2001ApJ...559..296D}.
The CAPRICE flight was in 1994, and quoted 600 MV  \citep{2000ApJ...532..653B}.
The AMS01 flight was in 1998 at the end of a solar minimum, and quoted $650\pm40$ MV  \citep{2002PhR...366..331A}
\footnote{The modulation levels quoted by the authors implicitly assume an interstellar CR spectrum, so the derivation may be a circular argument.}.
The Fermi-LAT data was taken in 2008-2010 during an extreme solar minimum of polarity opposite to that of 1994-5, 
but for which no reliable modulation level is available.
  The PAMELA data were taken during 2006--2010, and a modulation of 600 MV is used in their paper.

Fermi-LAT measured the electron spectrum from 7 GeV to 1 TeV, with unprecedented
accuracy. 
 Above 20 GeV the  Fermi-LAT electron spectrum can be fitted with a simple power law with spectral index 3.04. 
H.E.S.S. measured the electron spectrum from 340 GeV to 5 TeV with a tendency to be higher than
Fermi/LAT measurements in the region of overlap but in agreement within the systematics, and with a steepening
above 1 TeV.  ATIC~-~1~-~2 measured the  spectrum from 20 GeV to 2 TeV and is consistent
with Fermi measurements up to 300 GeV, but  found a peak between 300 and 700 GeV, which is not
confirmed by Fermi/LAT.
Energies above 100 GeV are not important for synchrotron (the corresponding frequencies lie beyond 1 THz ) but are included in the model and data for completeness. 

 The electron spectrum (1--625 GeV) recently measured by PAMELA \citep{2011PhRvL.106t1101A} is generally consistent with that measured by Fermi-LAT, although it is about 20\% higher in the 7-20 GeV range of energy overlap. The overall spectrum is also slightly softer than found by Fermi-LAT (and consistent with a rising positron fraction, see discussion in \cite{2011PhRvL.106t1101A}).
 We choose here to baseline on Fermi-LAT, while bearing in mind this difference, which does not affect our conclusions.
We do not address the high-energy positron excess reported by the PAMELA group \citet{2009Natur.458..607A}, since the main excess lies above the energies of importance for synchrotron, and in addition because the absolute PAMELA positron spectrum is not yet available.

 Lower energy (100 MeV and below) electron data are available from the Voyager \citep{2008JGRA..11311106W} 
  and Ulysses \citep{2005AdSpR..35..605H} spacecraft, but they are not constrained by synchrotron and will not be used here.

\section{Results}
\subsection{Pure Diffusion Model}

We consider first a `pure diffusion' model, with a halo height of  4 kpc.
The complete set of GALPROP parameters are given in \citet{2041-8205-722-1-L58}, model z04LMPDS.
 The electron injection spectrum breaks at 4 GeV and 50 GeV, with indices 1.6/2.5/2.2.
The break at 50 GeV is to fit the Fermi-LAT low-energy upturn, the break at 4 GeV to fit low-frequency synchrotron.  A cutoff at 2 TeV is introduced to reproduce the H.E.S.S. data, although this has no effect for the synchrotron, the corresponding frequencies being far too high.
Fig~\ref{PD_lepton_spectra_basic} shows the interstellar electron and positron spectra for this model, and also for various modulation levels using the force-field approximation.
Fig~\ref{PD_sync_spectra_basic} shows that this model gives a reasonable fit to the synchrotron spectrum, and illustrates the role played by the secondary leptons which have a steeper spectrum than primaries and contribute significantly to the low-frequency synchrotron. Fig~\ref{PD_sync_indices_basic} shows the spectral index as a function of frequency for this model, for primary and secondary leptons and for total leptons.

\begin{figure}
\includegraphics[width=0.35\textwidth, angle=0]{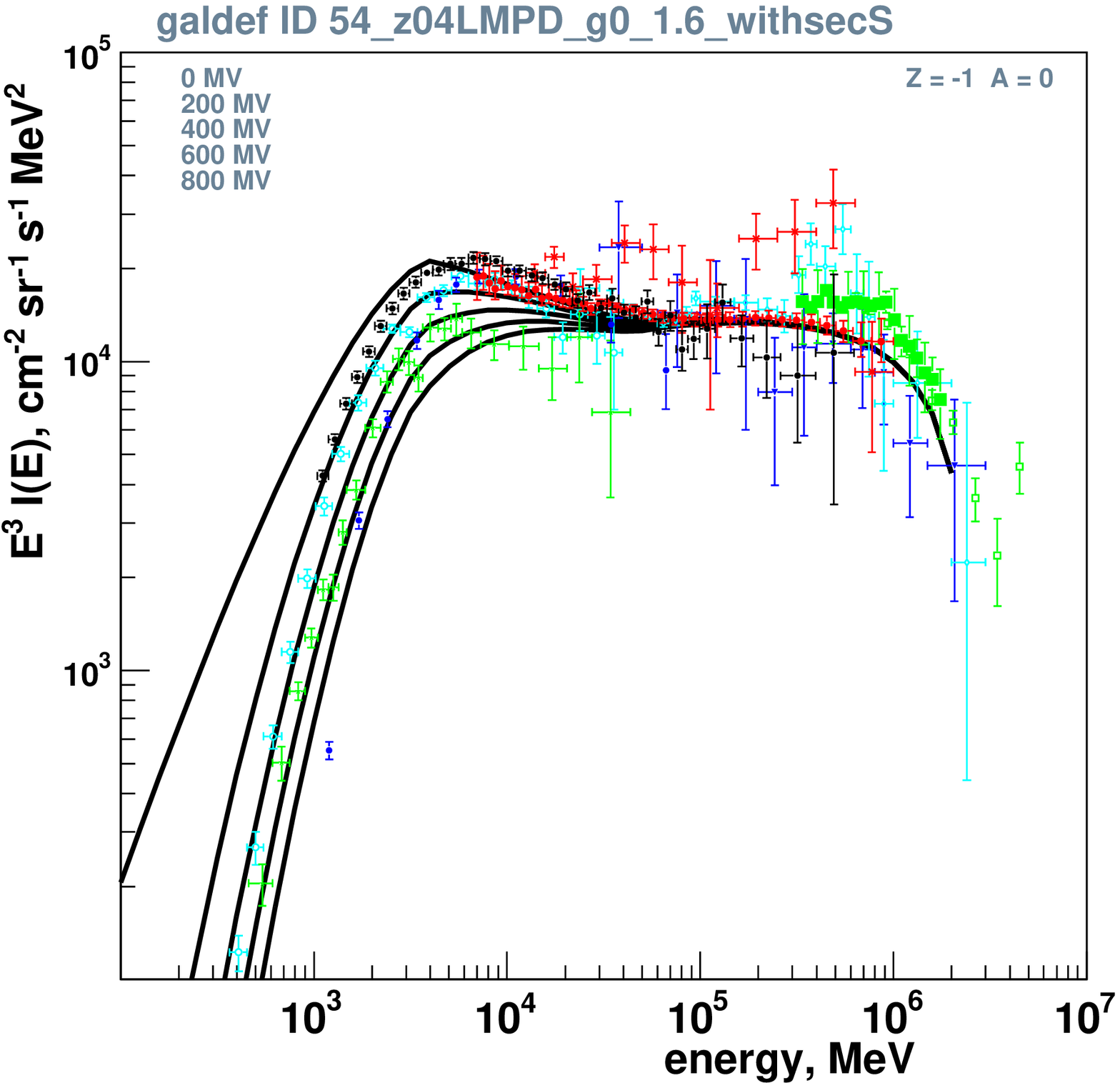}
\includegraphics[width=0.35\textwidth, angle=0]{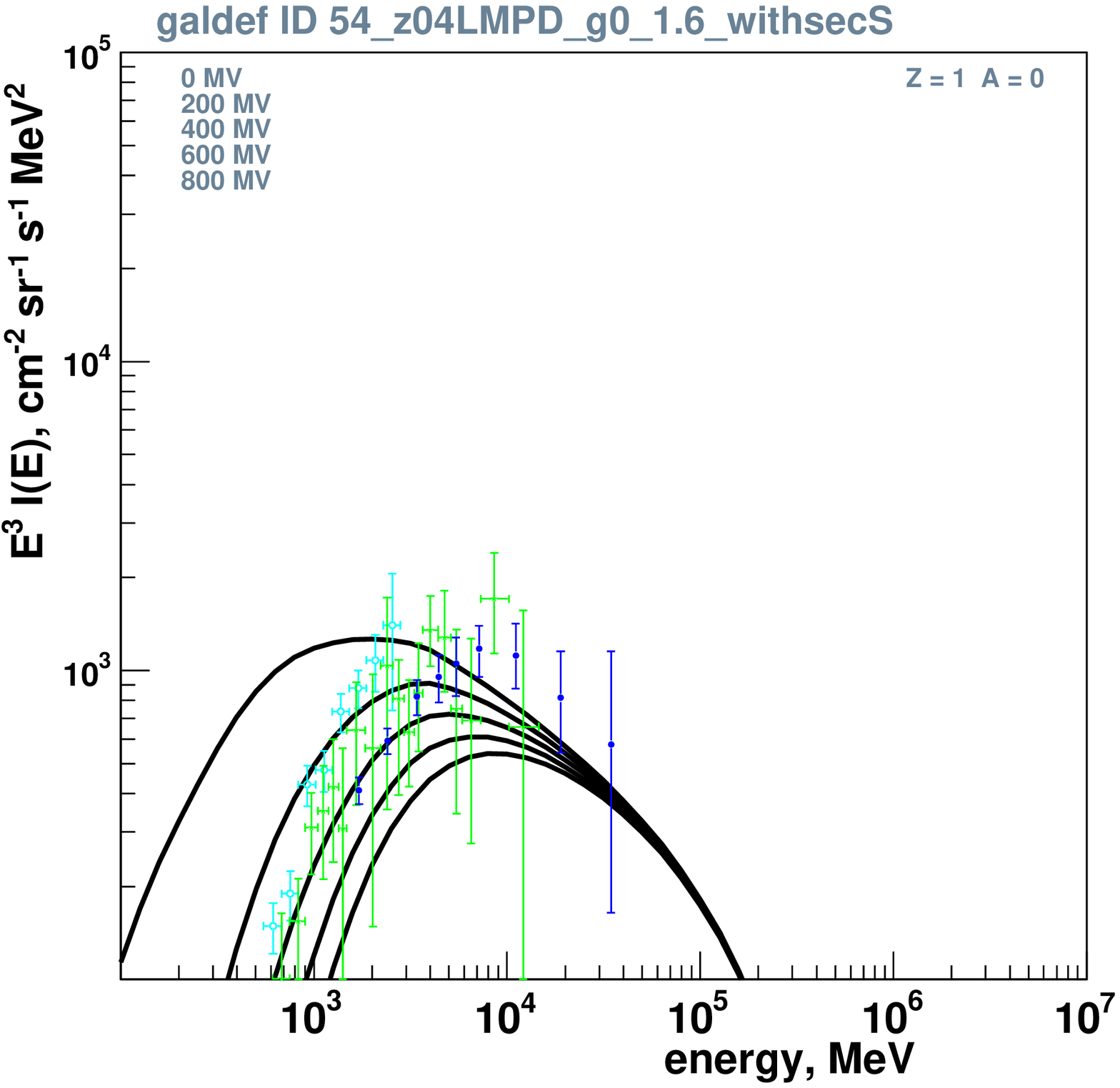}
\caption{Electron (upper) and positron (lower) spectra for pure diffusion model with primary low-energy electron injection index 1.6.  Modulation $\Phi$=0,200,400,600,800 MV.   NB Fermi-LAT includes positrons.
Cyan open circles: AMS01; green crosses and filled circles: CAPRICE; blue squares: HEAT; red filled circles: Fermi-LAT; black filled circles: PAMELA; blue triangles: SANRIKU; red crosses: BETS, PPT-BETS; cyan open circles: ATIC-1-2; green filled and open squares: H.E.S.S. For references see text.
}
\label{PD_lepton_spectra_basic}
\end{figure}


\begin{figure}
\includegraphics[width=0.35\textwidth, angle=0]{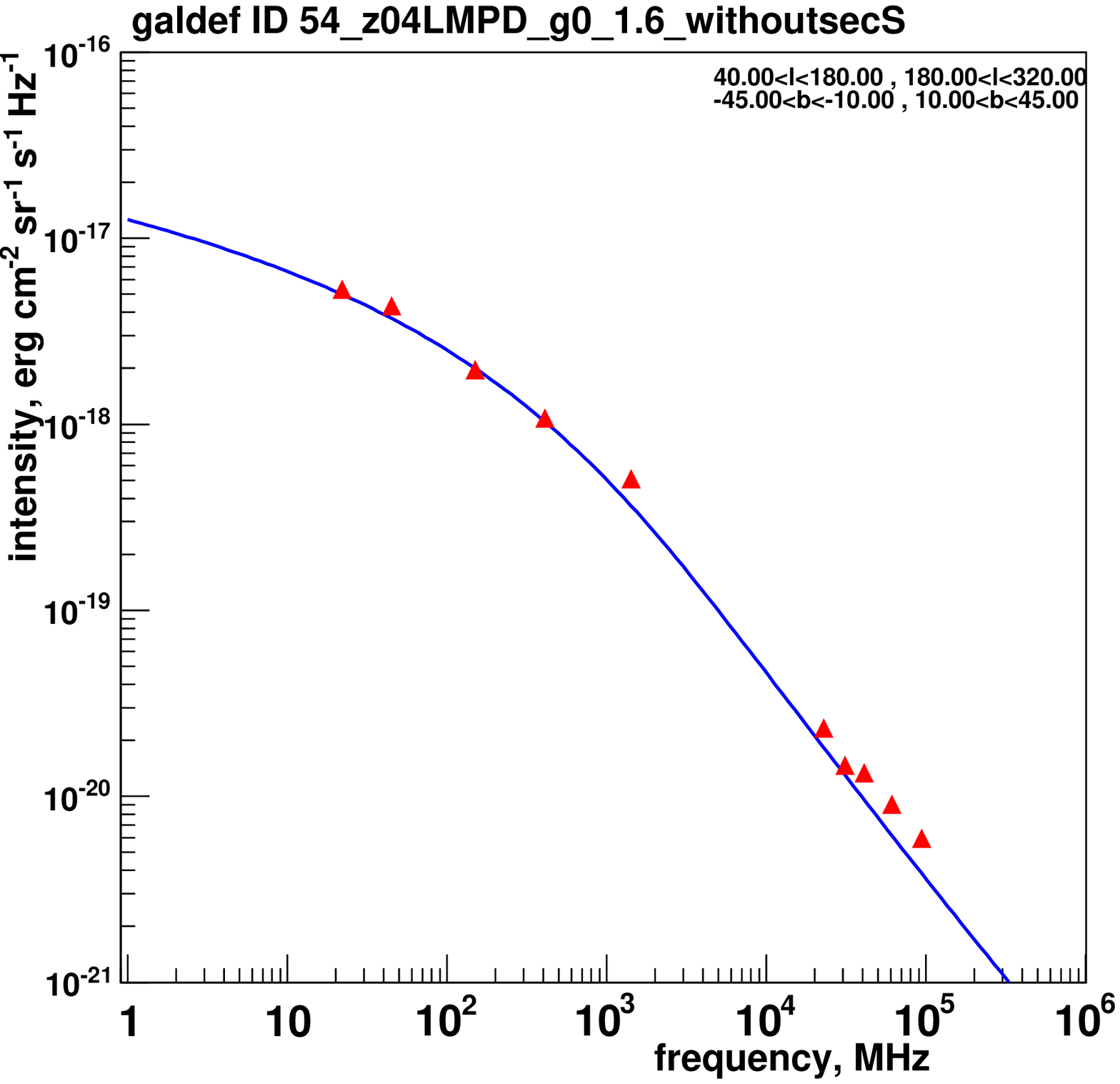}
\includegraphics[width=0.35\textwidth, angle=0]{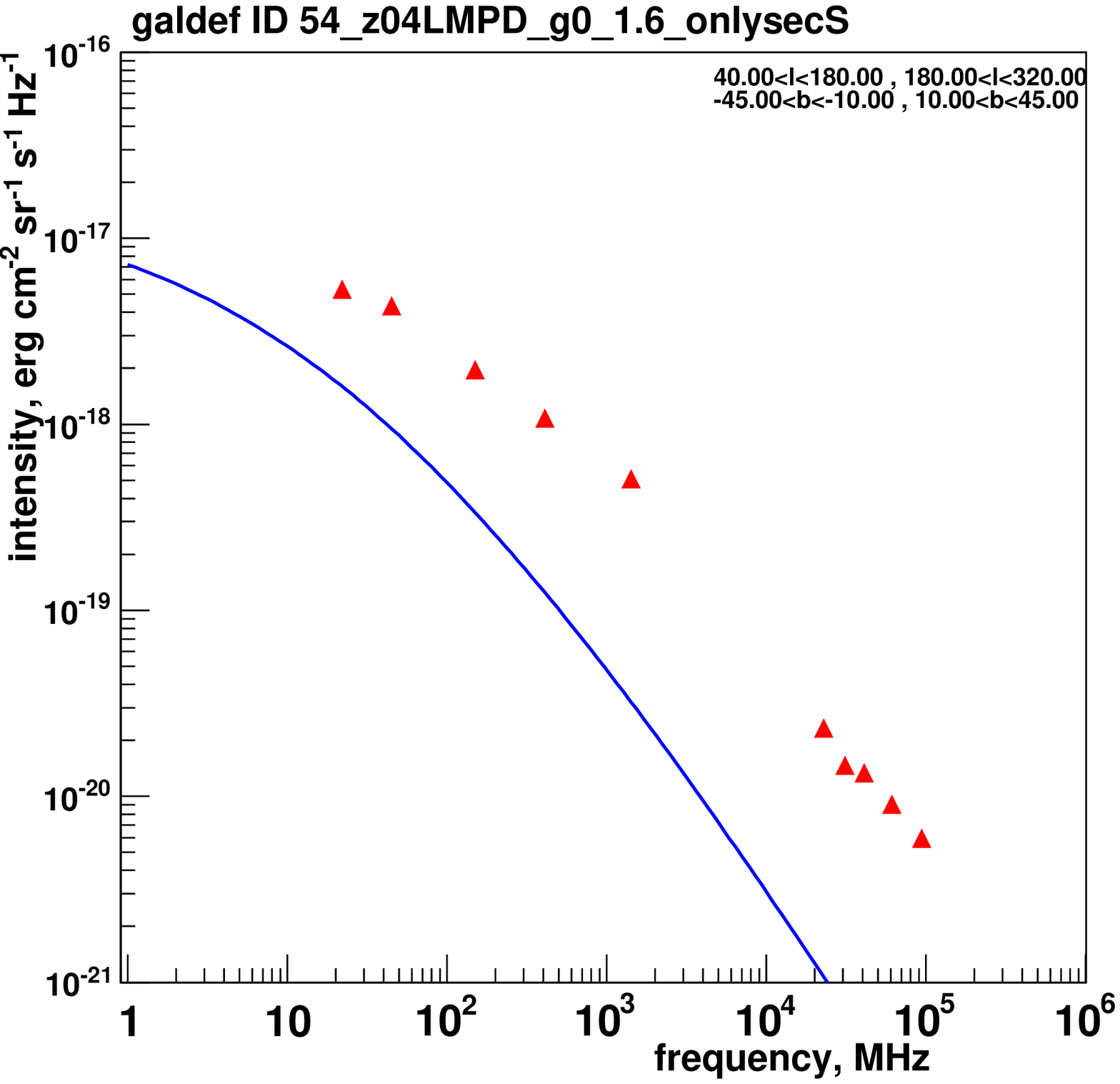}
\includegraphics[width=0.35\textwidth, angle=0]{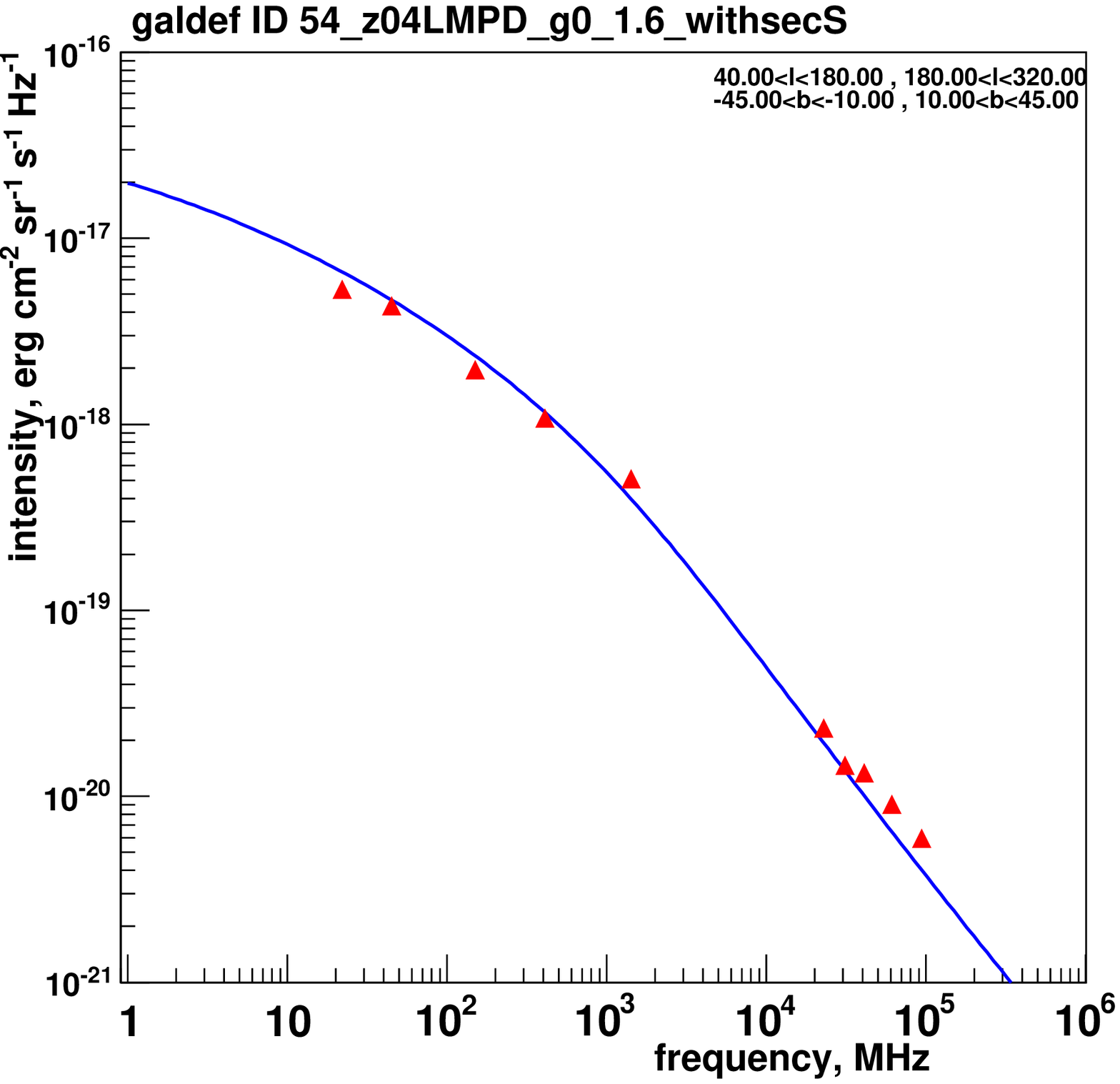}
\caption{Synchrotron spectra for pure diffusion model with primary low-energy electron injection index 1.6. Synchrotron from primary electrons  (upper), secondary leptons (middle) and total (lower). For synchrotron data references see Section 4.2. 
 }
\label{PD_sync_spectra_basic}
\end{figure}



\begin{figure}
\includegraphics[width=0.34\textwidth, angle=0]{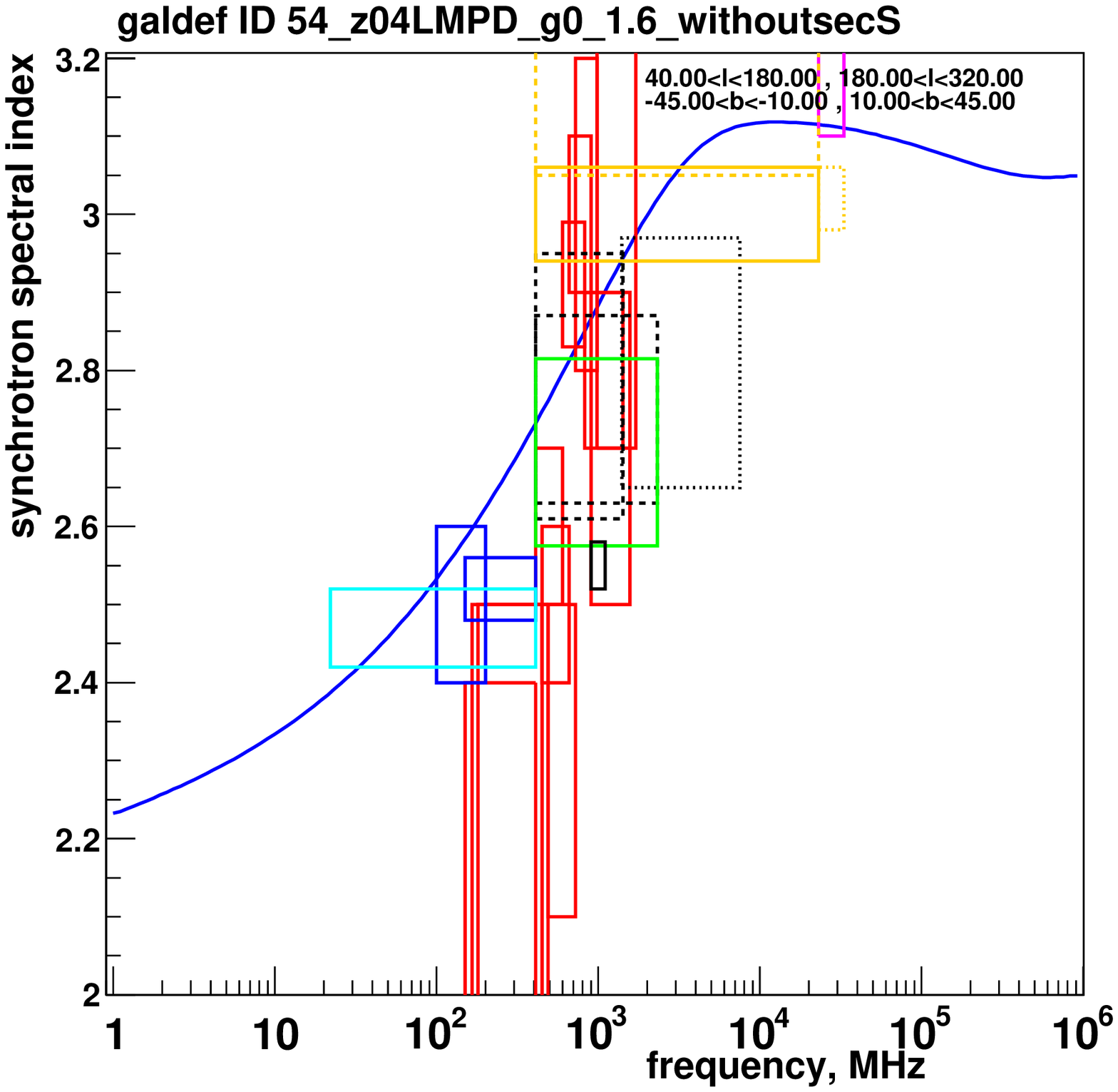}
\includegraphics[width=0.34\textwidth, angle=0]{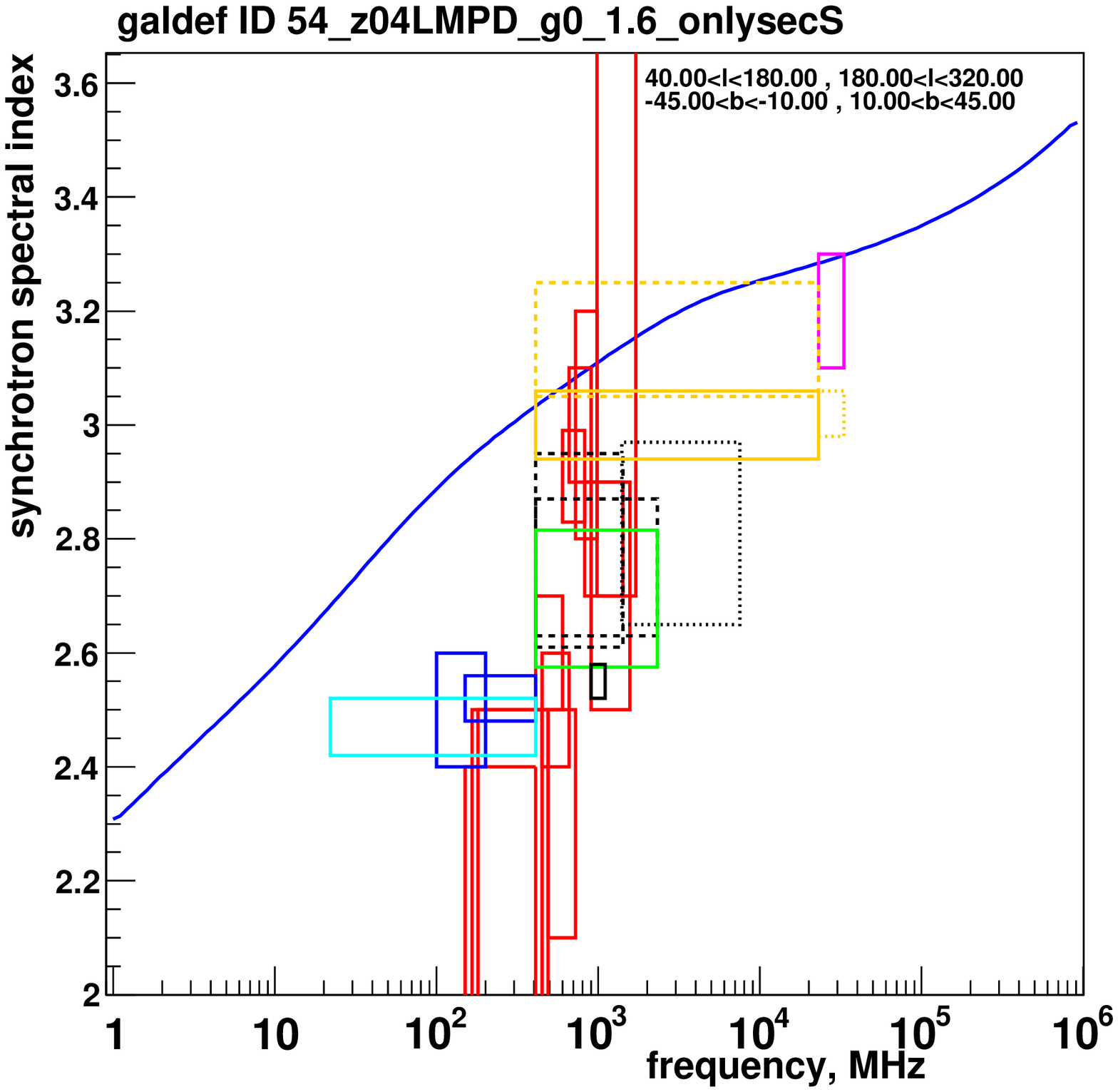}
\includegraphics[width=0.34\textwidth, angle=0]{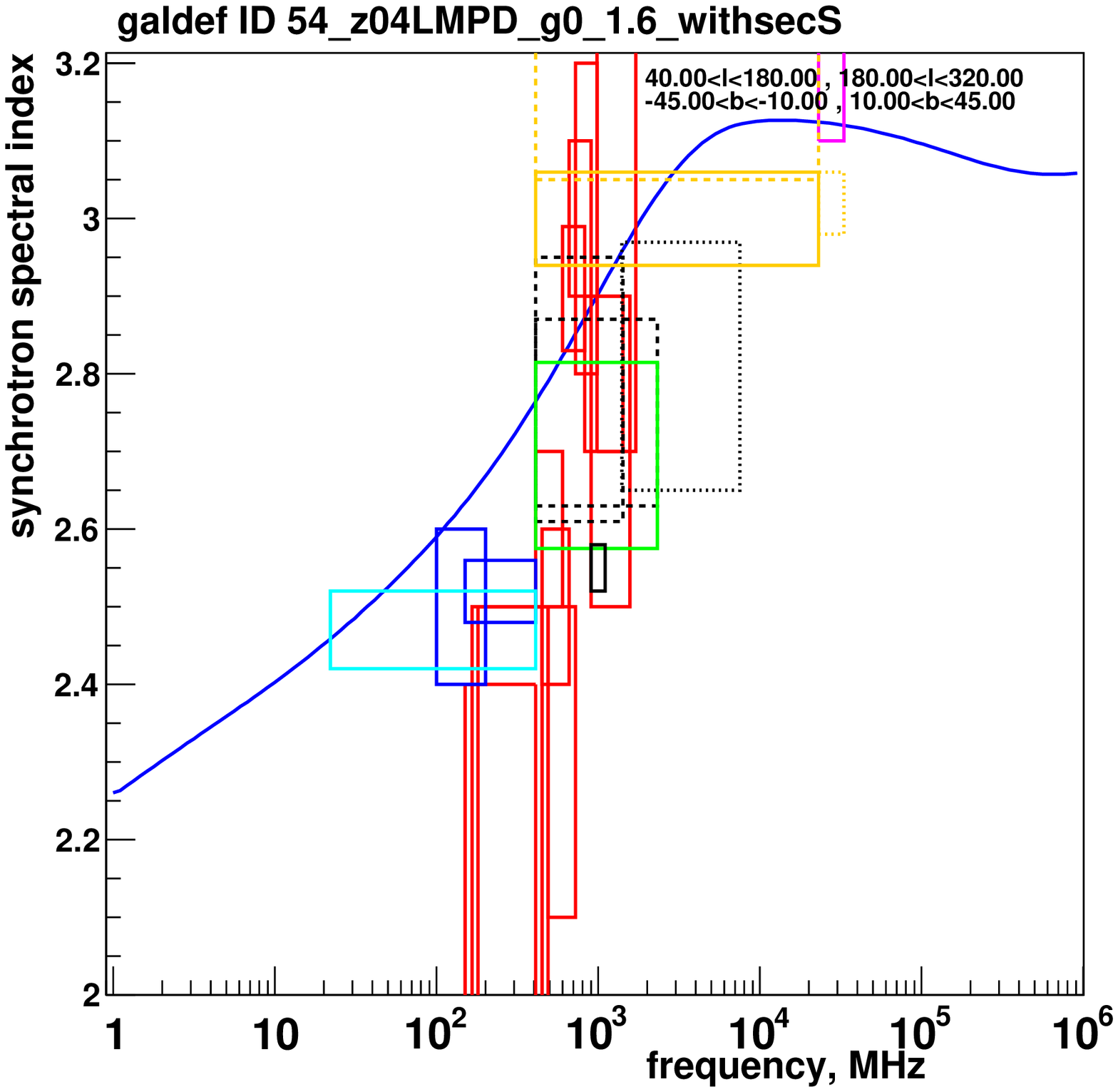}
\caption{ Synchrotron spectral index for pure diffusion model with primary low-energy electron injection index 1.6. Synchrotron from primary electrons  (upper), secondary leptons (middle) and total (lower). The experimental ranges are based on values from the literature as reviewed in the text, and are only intended to be indicative of the general trend since the  measurements cover different sky areas.
 Data:
 red:          \citet{2008ApJ...688...32T};  
 blue:         \citet{2008AJ....136..641R};  
 cyan:         \citet{1999A&AS..137....7R};  
 black dashed: \citet{2002A&A...387...82G};  
 black dotted: \citet{1998ApJ...505..473P};  
 green:        \citet{2003A&A...410..847P};  
 cyan:         \citet{2007ApJ...665..355K};  
 orange full  :\citet{2008A&A...490.1093M};  
 orange dashed:\citet{2009ApJS..180..265G};  
 orange dotted:\citet{2009ApJ...701.1804D};  
 black  full  :\citet{2011ApJ...734....4K}.  
}
\label{PD_sync_indices_basic}
\end{figure}

We next show the effect of varying the low-energy ($<$4 GeV) electron injection index, from 1.0 to 2.5. 
Fig~\ref{PD_electron_spectra_various} shows the interstellar electron spectra for these models, and also for various modulation levels using the force-field approximation. It is clear that the electron data alone cannot distinguish the models due to the modulation uncertainty, so that the synchrotron constraints are essential.  

\begin{figure}
\includegraphics[width=0.23\textwidth, angle=0]{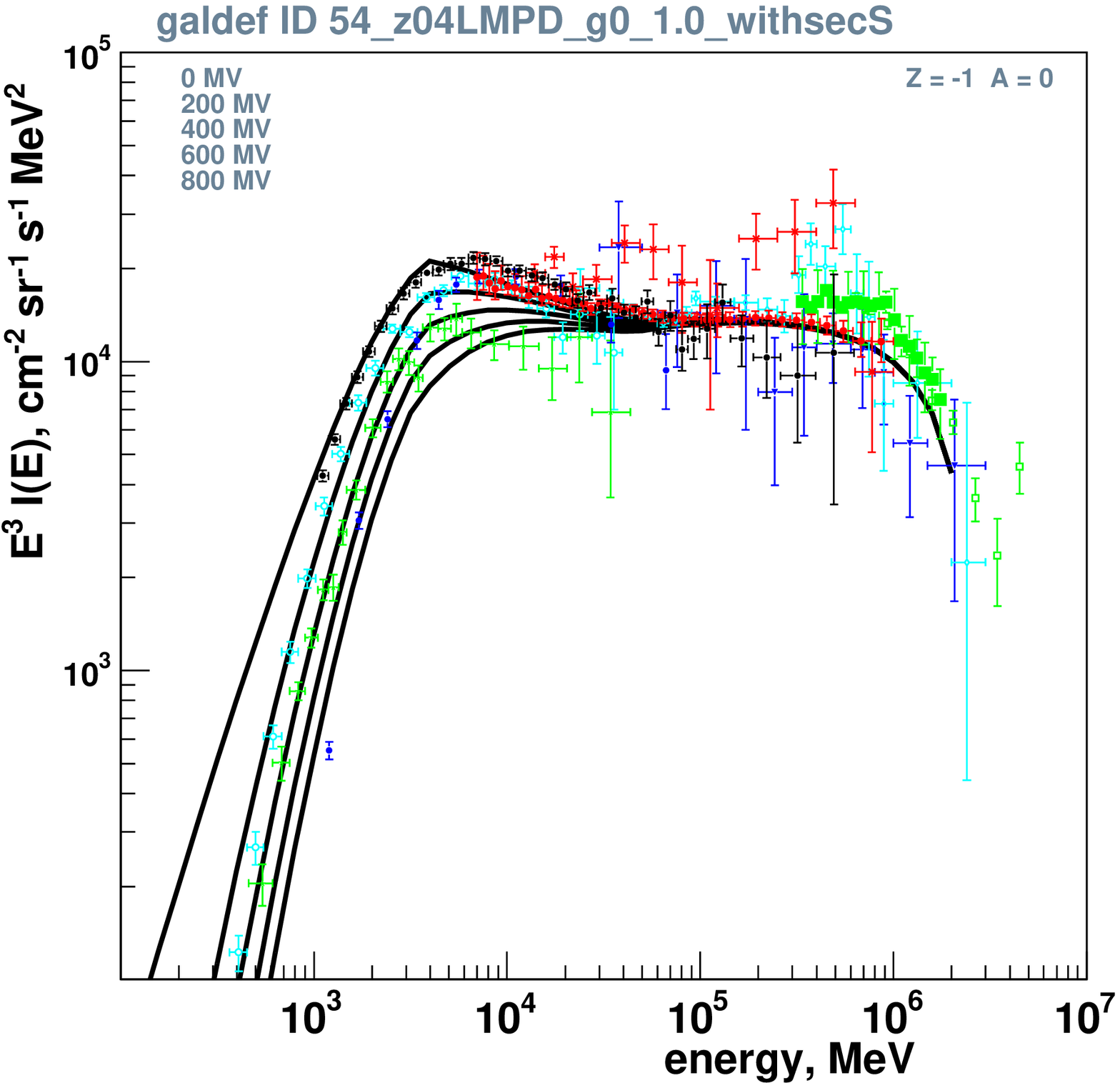}
\includegraphics[width=0.23\textwidth, angle=0]{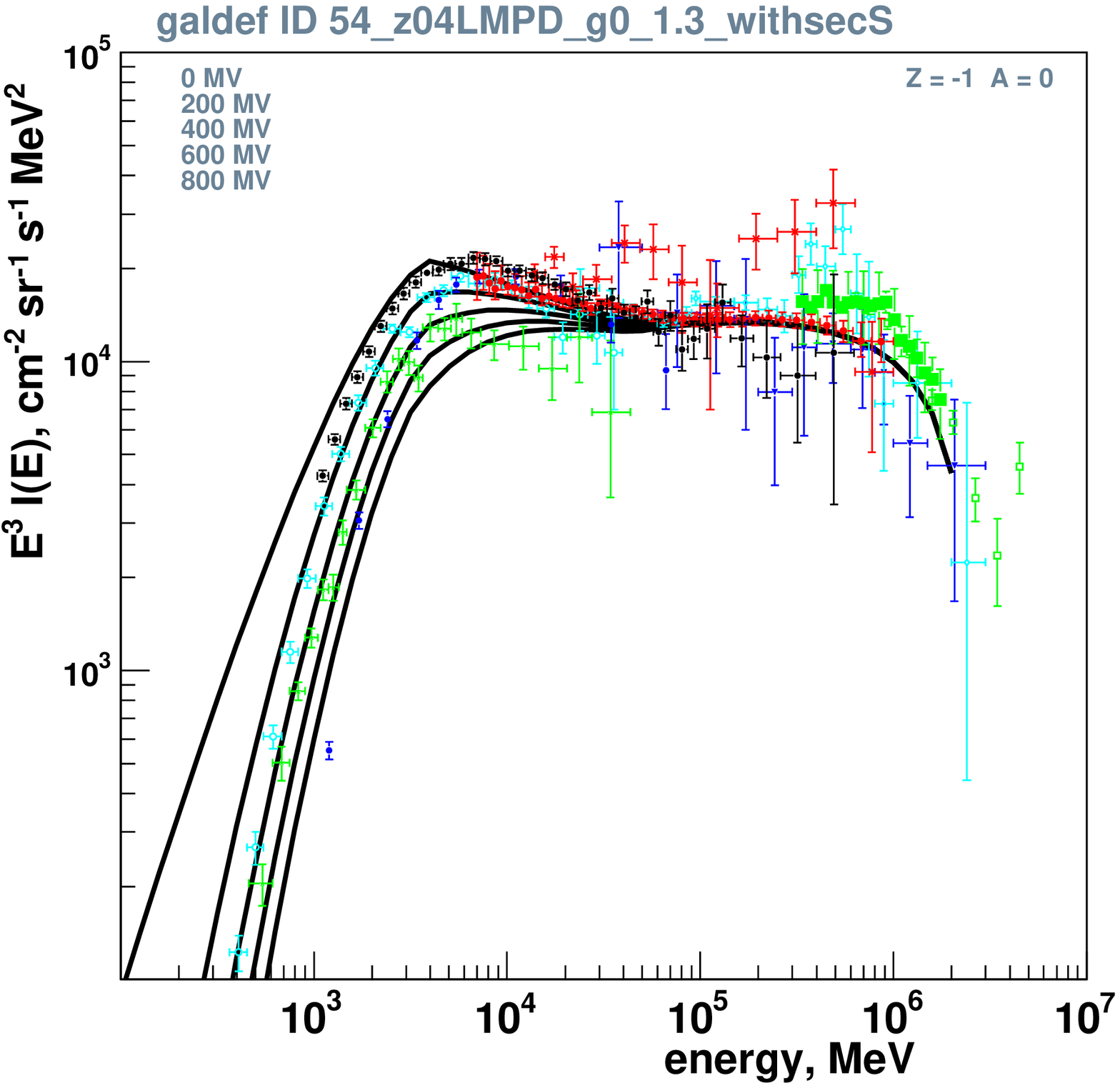}
\includegraphics[width=0.23\textwidth, angle=0]{gcr_spectra_54_z04LMPD_g0_1.6_withsecS_Z-1_A0.eps}
\includegraphics[width=0.23\textwidth, angle=0]{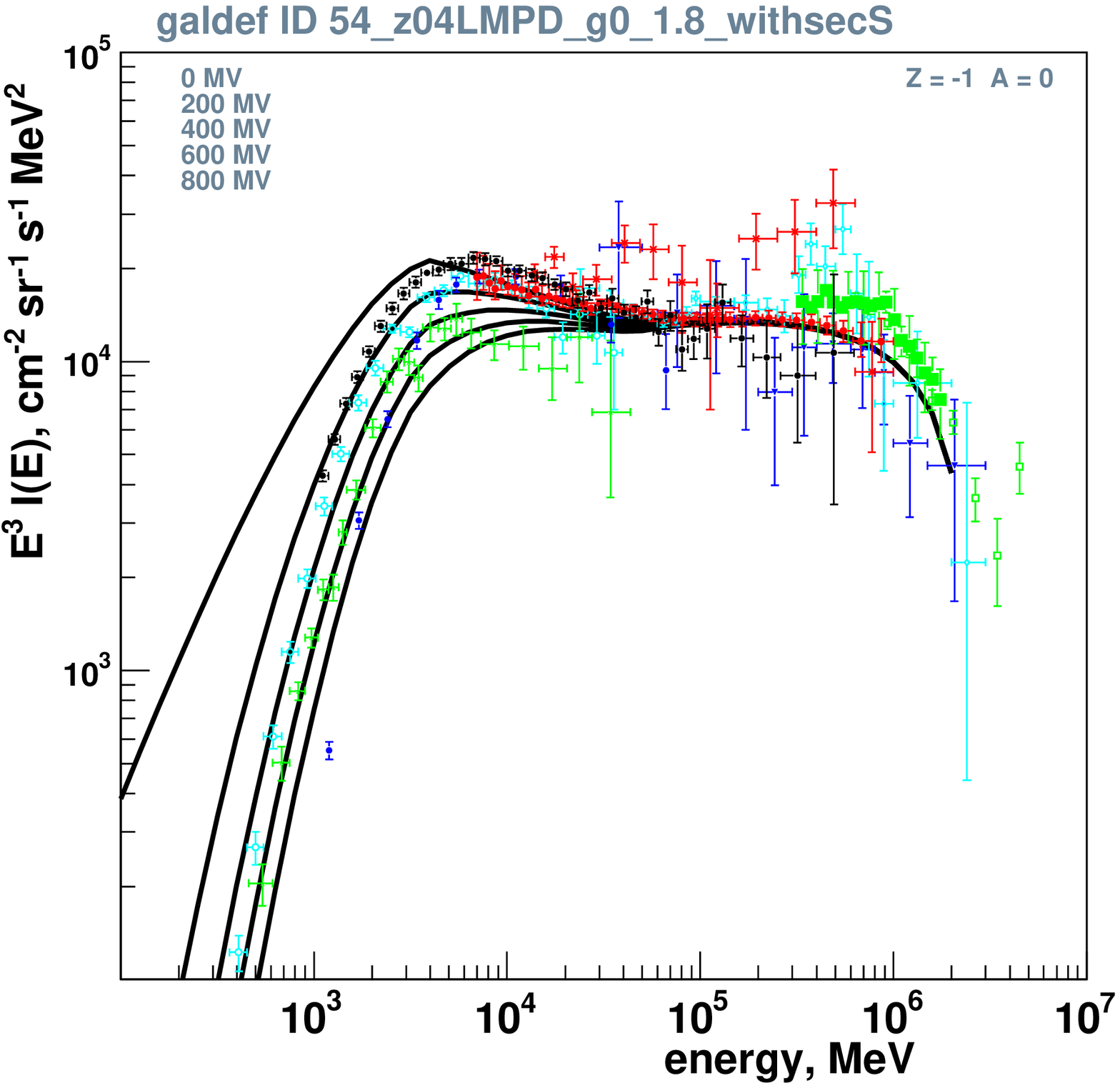}
\includegraphics[width=0.23\textwidth, angle=0]{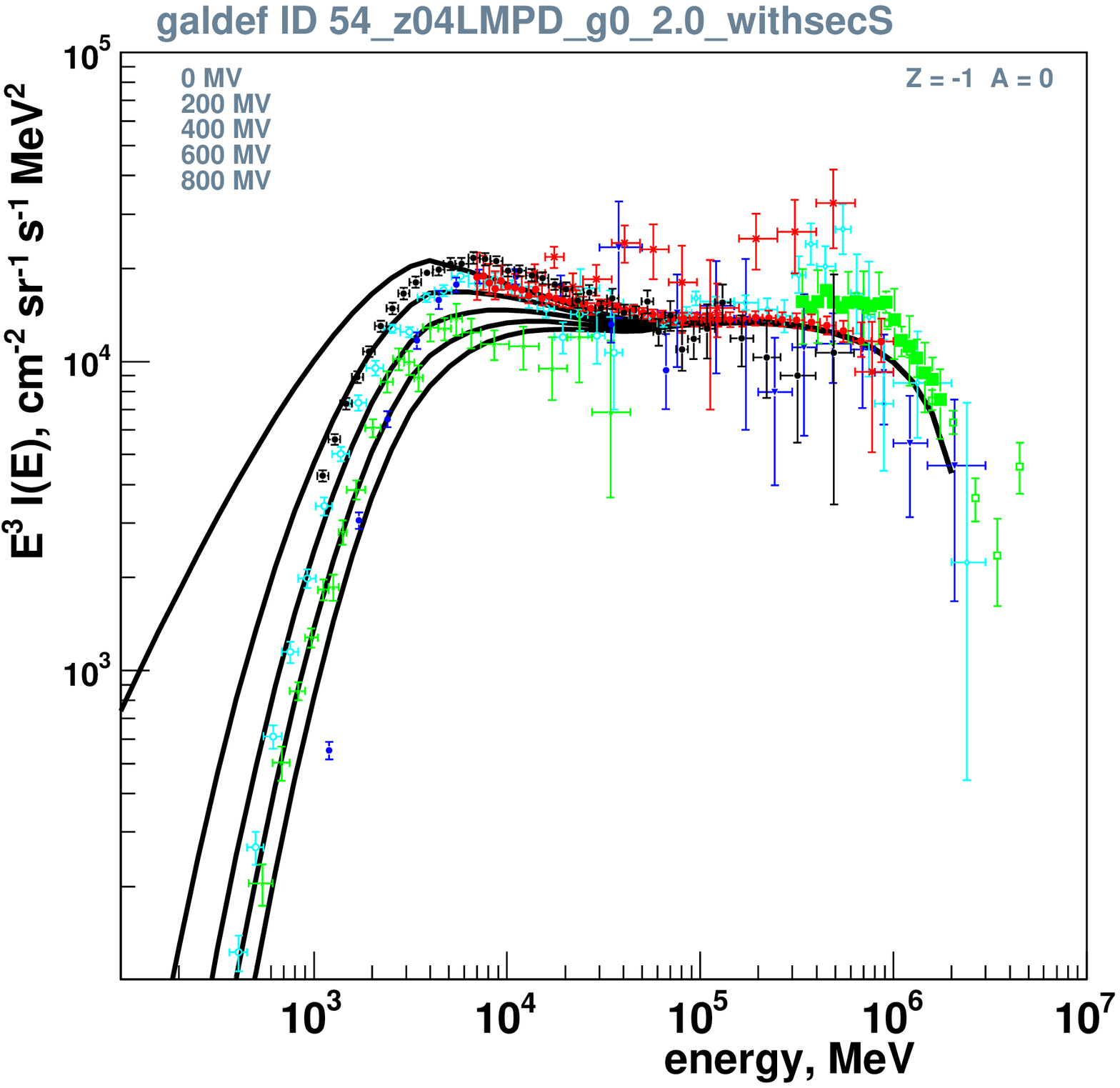}
\includegraphics[width=0.23\textwidth, angle=0]{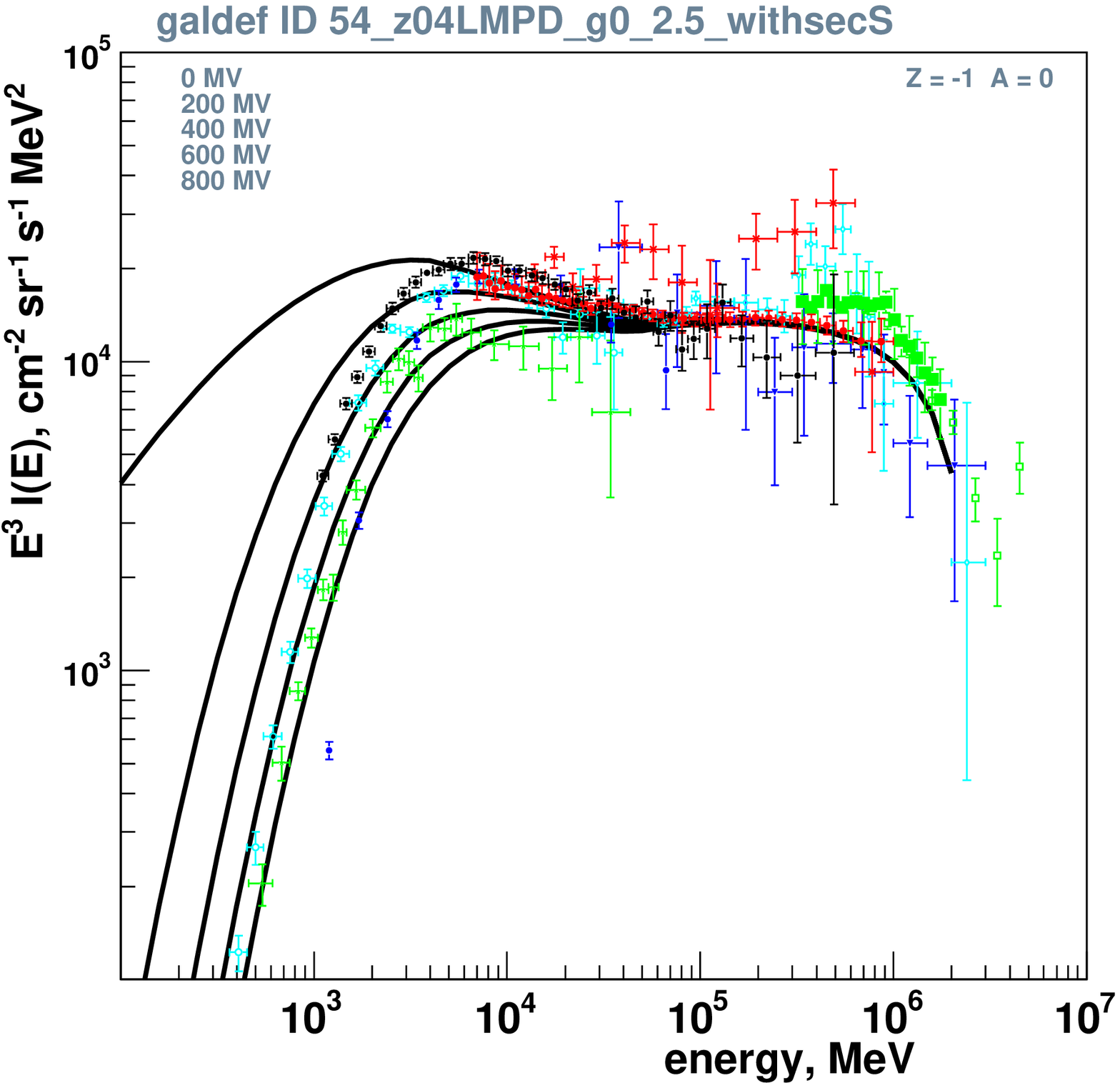}
\caption{Electron spectra for pure diffusion model, low-energy electron injection index 1.0,1.3,1.6, 1.8,2.0,2.5.  Modulation $\Phi$=0,200,400,600,800 MV. Data as in figure~\ref{PD_lepton_spectra_basic}.  }
\label{PD_electron_spectra_various}
\end{figure}

Fig~\ref{PD_sync_spectra_various} shows that a low-energy primary electron injection index of 2.0 is at the limit of the low-frequency synchrotron data. The best fit is actually  for an injection index around 1.3. Fig~\ref{PD_sync_indices_various} shows the synchrotron spectral index for these models, compared to values from the literature described in Section 4.1.

\begin{figure}
\centering
\includegraphics[width=0.23\textwidth, angle=0]{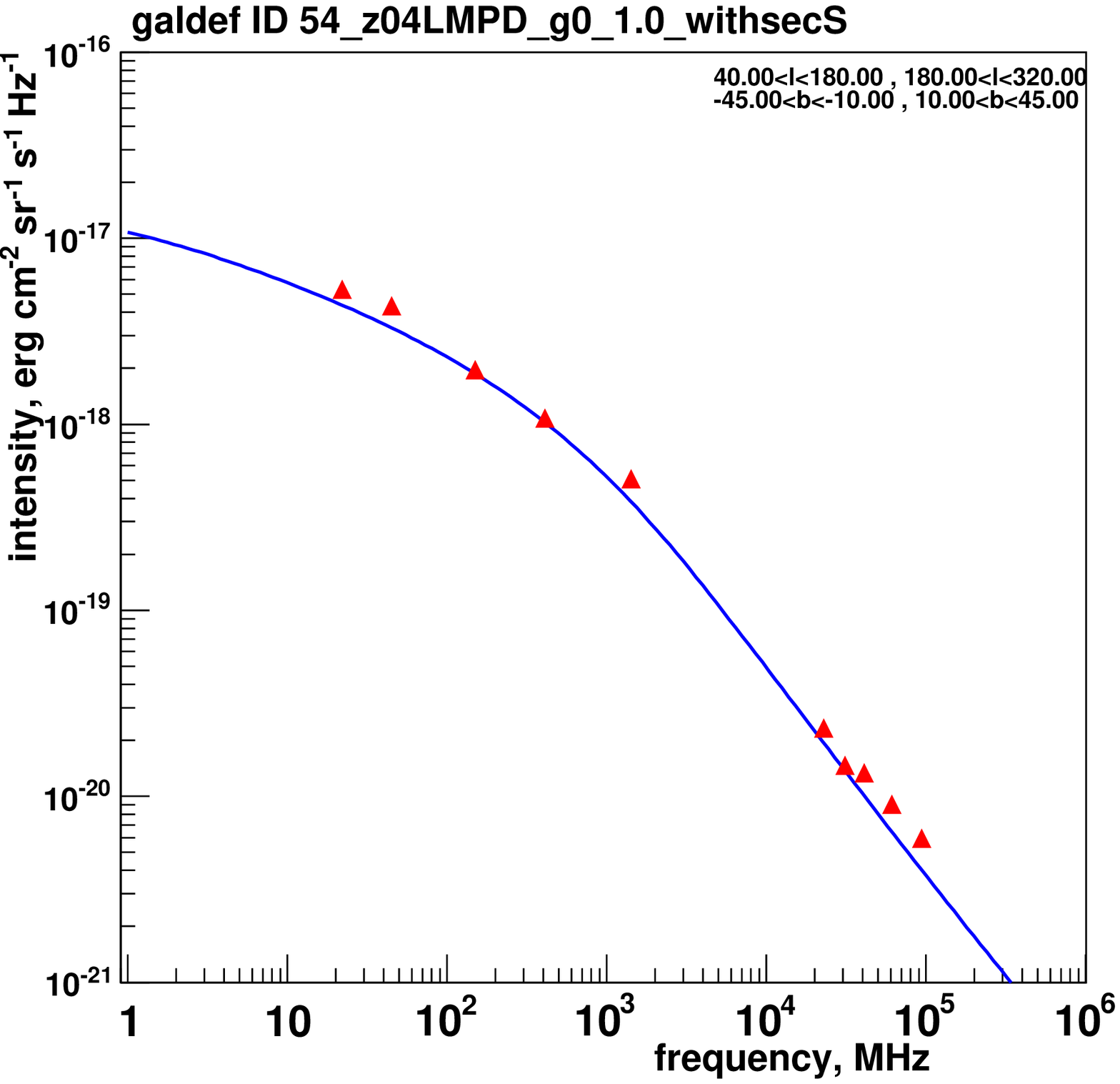}
\includegraphics[width=0.23\textwidth, angle=0]{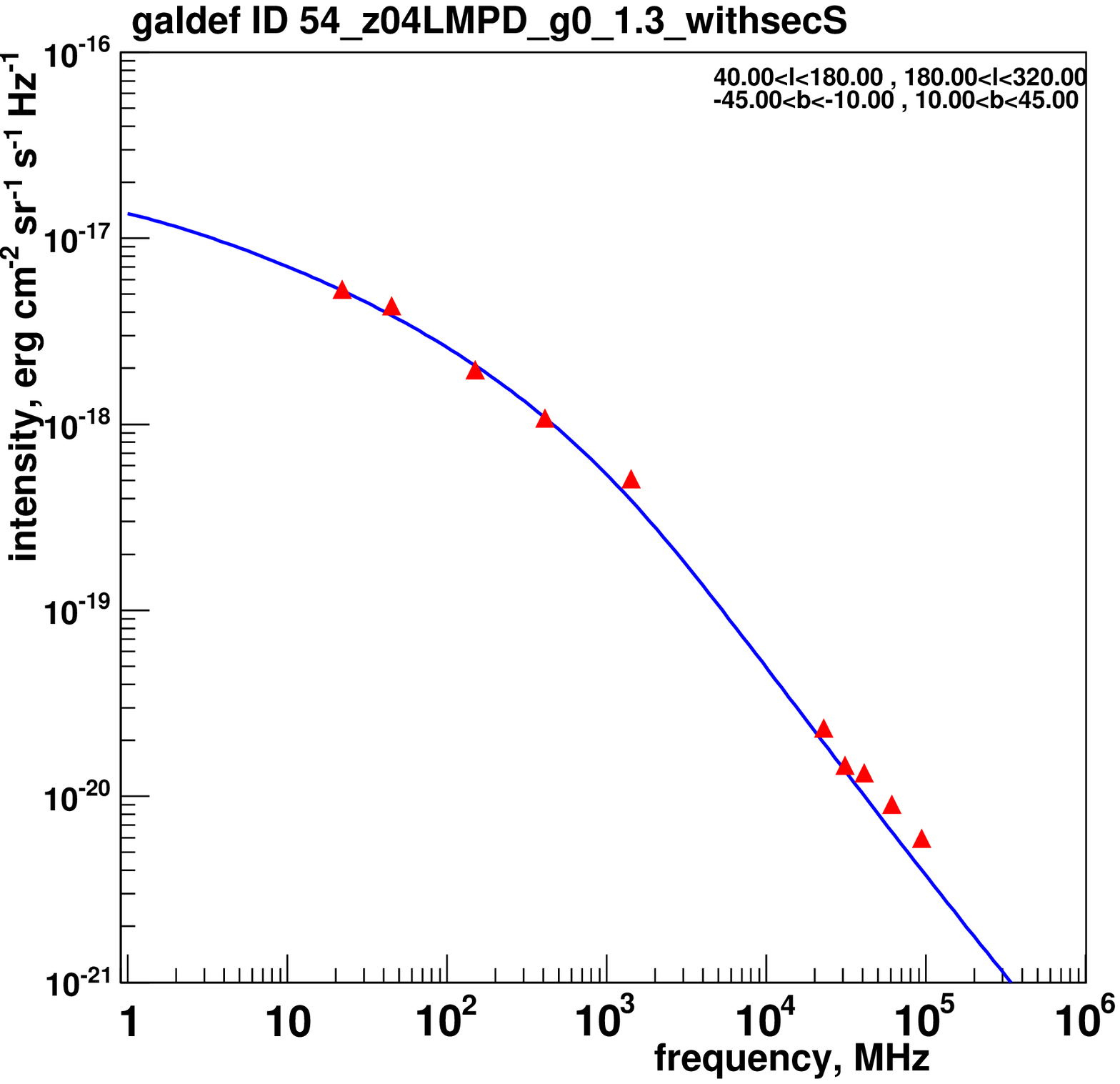}
\includegraphics[width=0.23\textwidth, angle=0]{synchrotron_spectrum_54_z04LMPD_g0_1.6_withsecS_l_40.00_180.00_180.00_320.00_b_-45.00_-10.00_10.00_45.00_fig_reg_1kpc_ranB7.5_R30_z4kpc.eps}
\includegraphics[width=0.23\textwidth, angle=0]{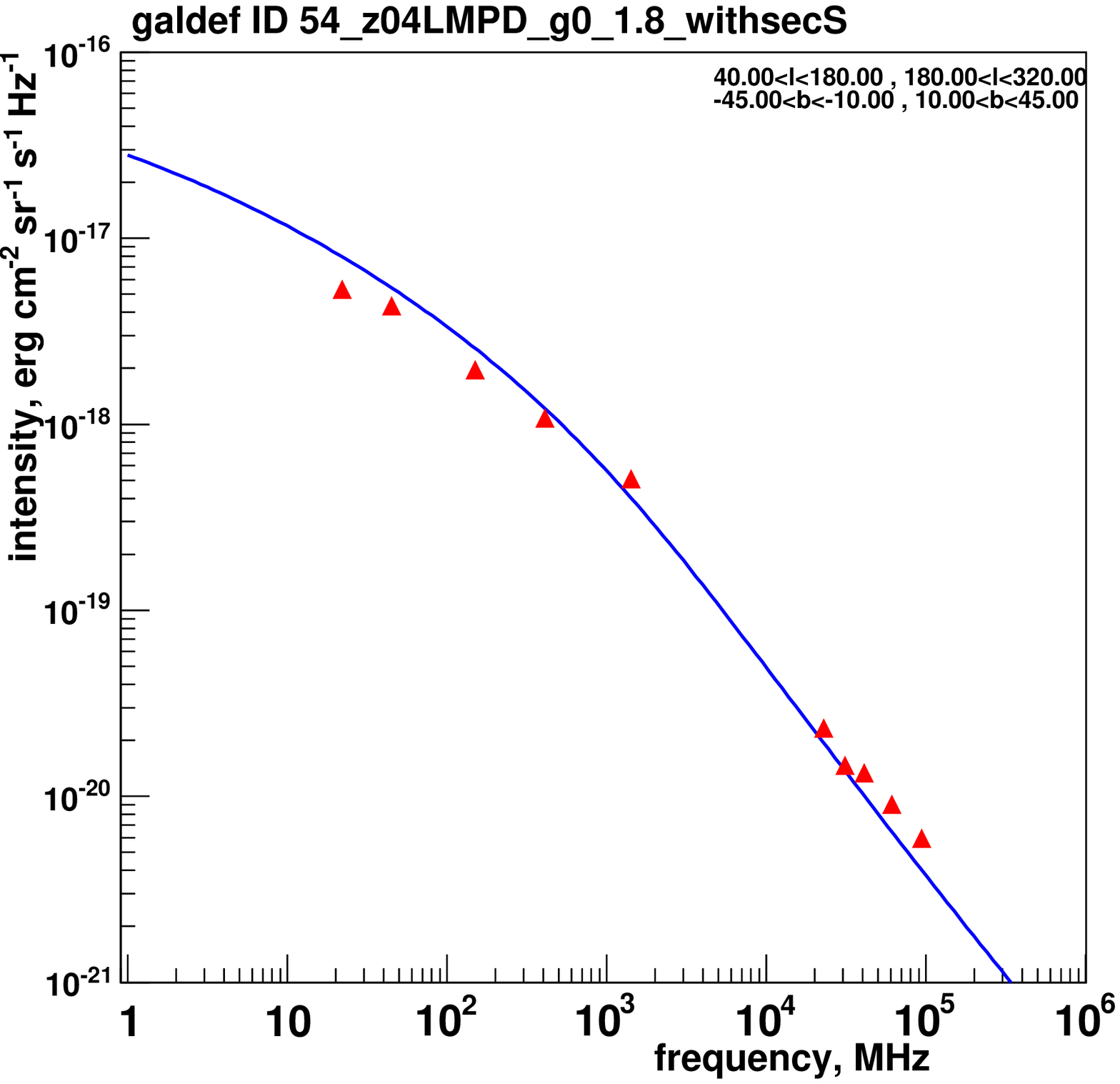}
\includegraphics[width=0.23\textwidth, angle=0]{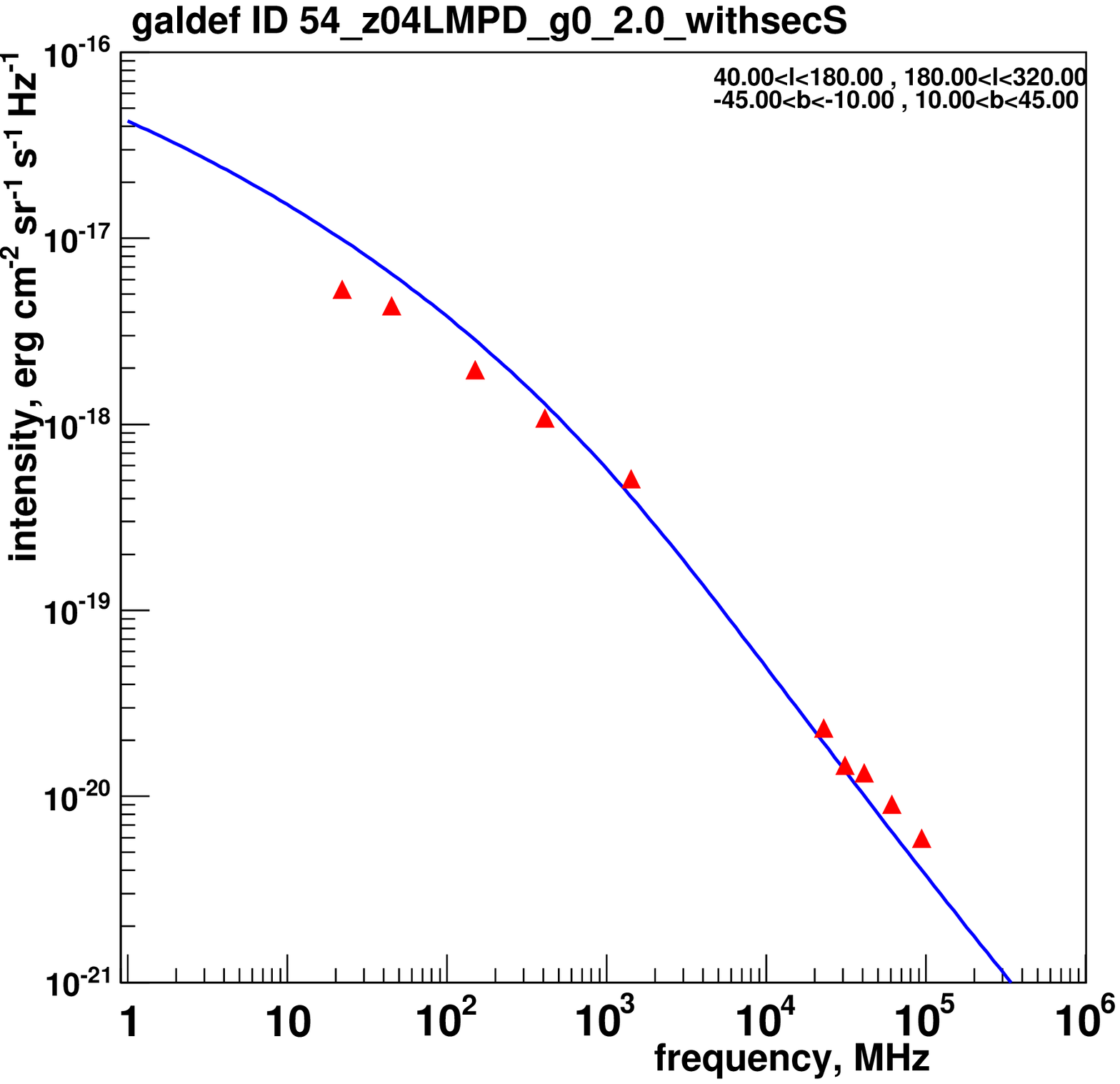}
\includegraphics[width=0.23\textwidth, angle=0]{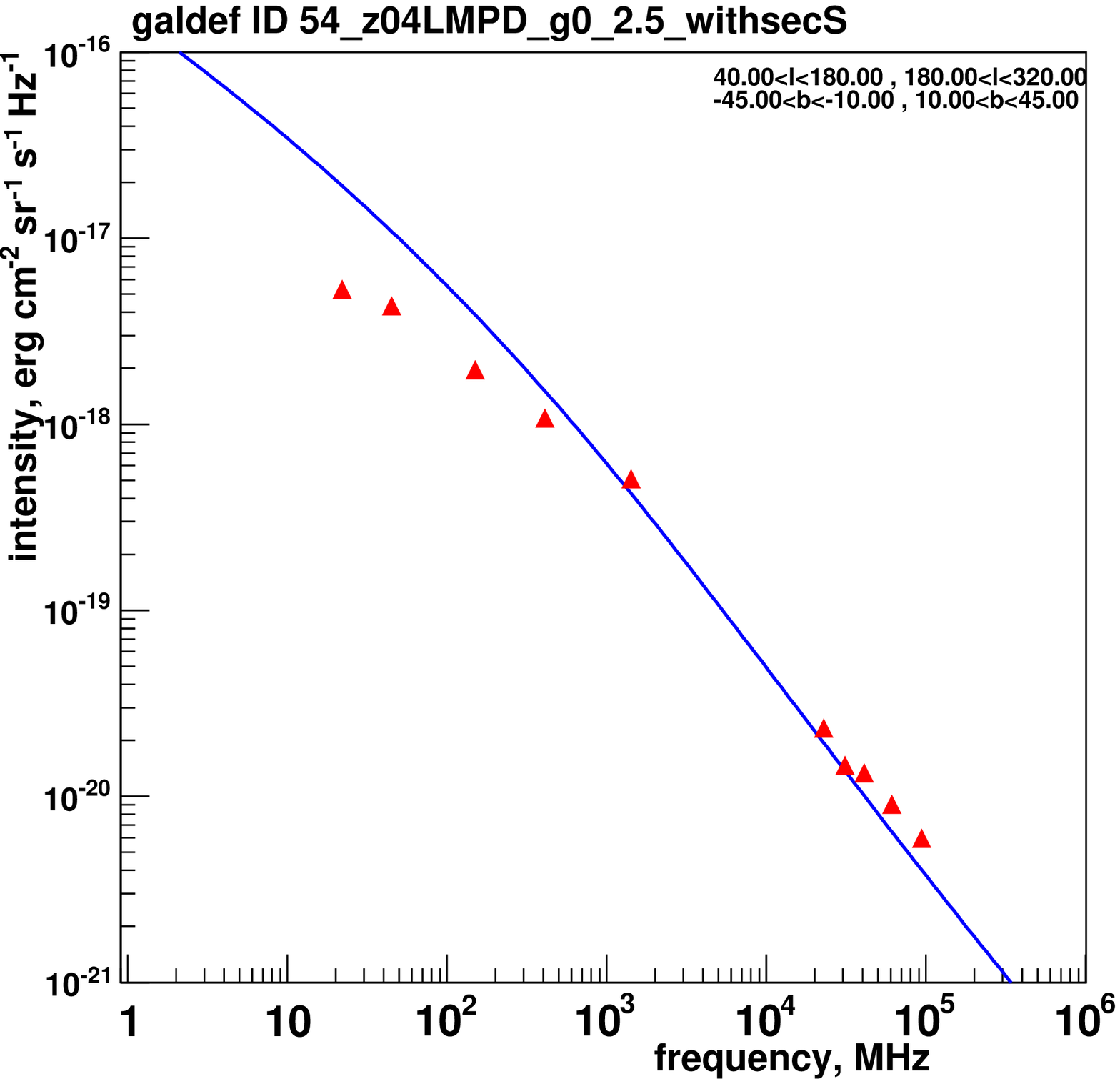}
\caption{Synchrotron spectra for pure diffusion model with low-energy electron injection index (left to right, top to bottom) 1.0, 1.3, 1.6, 1.8, 2.0, 2.5. Including secondary leptons. Data as in figure~\ref{PD_sync_spectra_basic}. }
\label{PD_sync_spectra_various}
\end{figure}

\begin{figure}
\centering
\includegraphics[width=0.23\textwidth, angle=0]{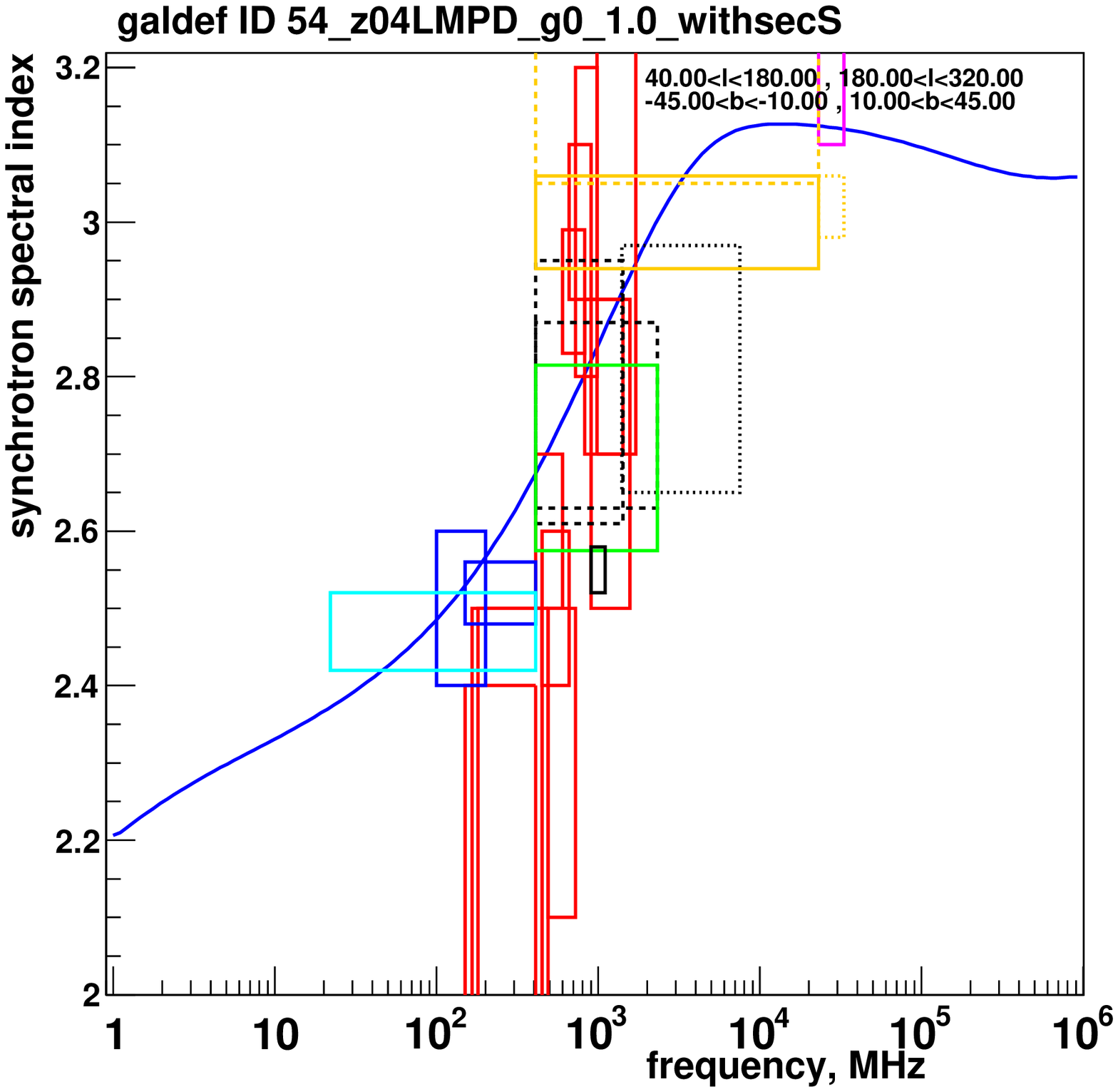}
\includegraphics[width=0.23\textwidth, angle=0]{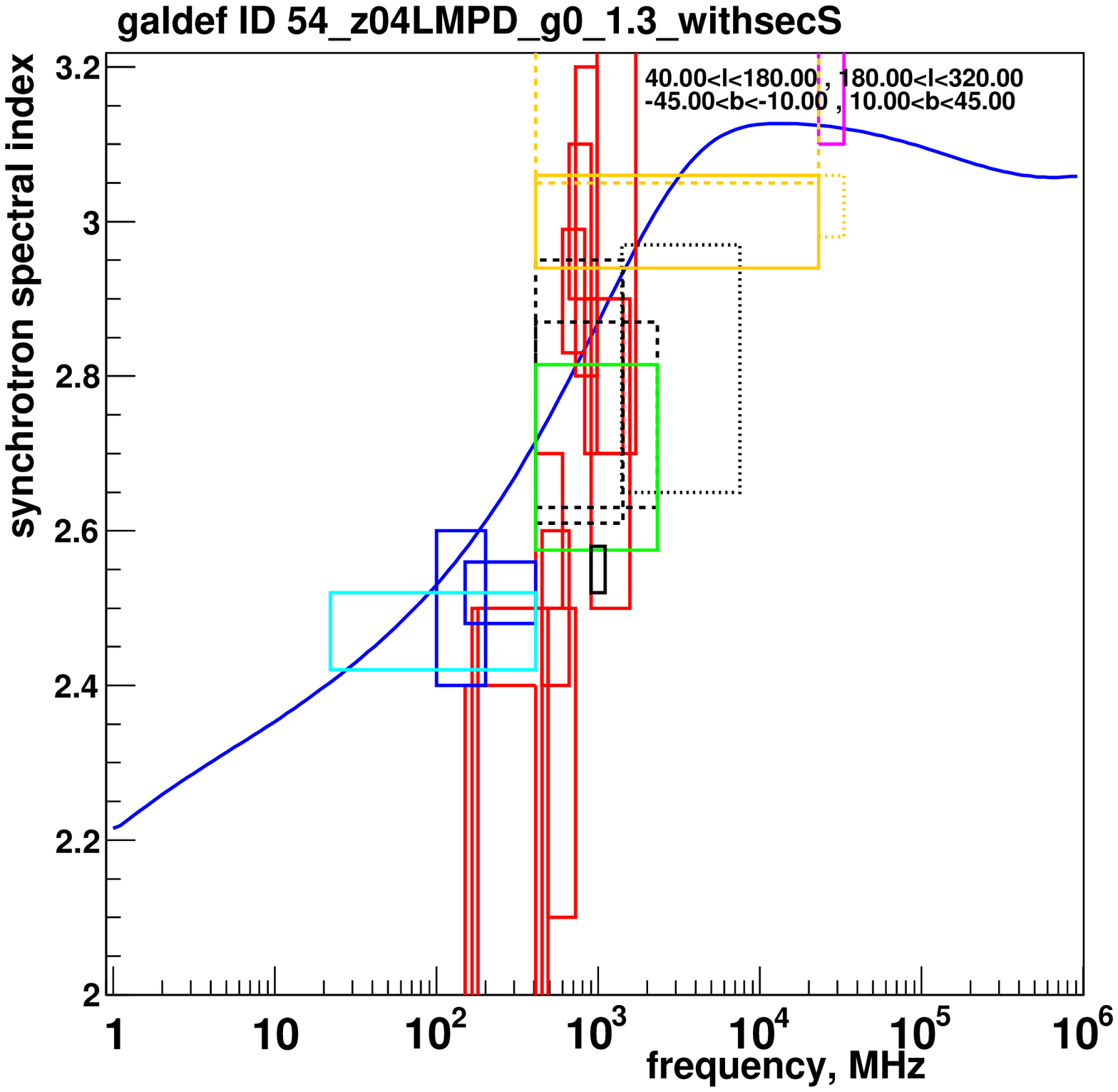}
\includegraphics[width=0.23\textwidth, angle=0]{synchrotron_spectral_index_54_z04LMPD_g0_1.6_withsecS_l_40.00_180.00_180.00_320.00_b_-45.00_-10.00_10.00_45.00_fig_reg_1kpc_ranB7.5_R30_z4kpc.eps}
\includegraphics[width=0.23\textwidth, angle=0]{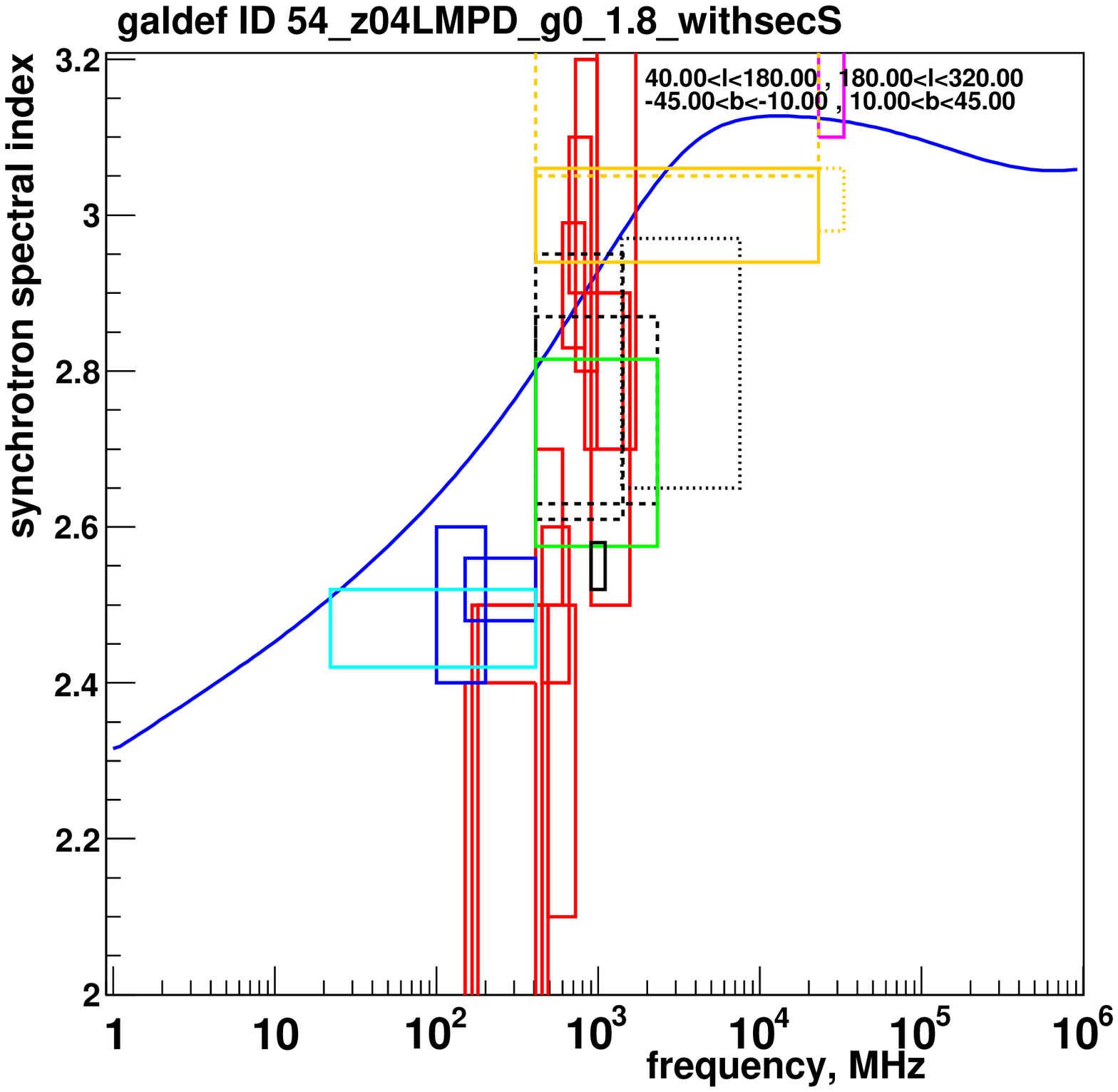}
\includegraphics[width=0.23\textwidth, angle=0]{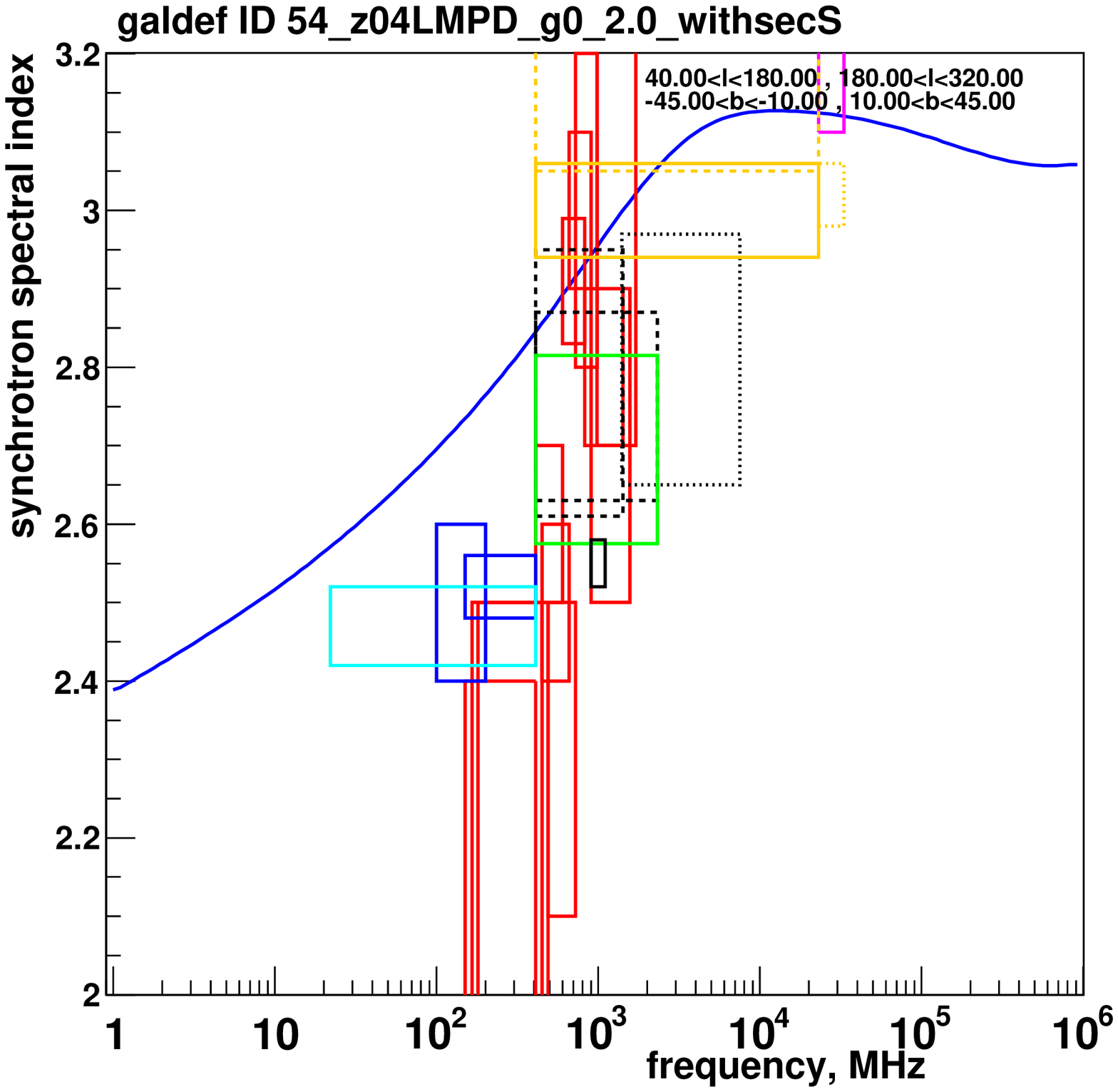}
\includegraphics[width=0.23\textwidth, angle=0]{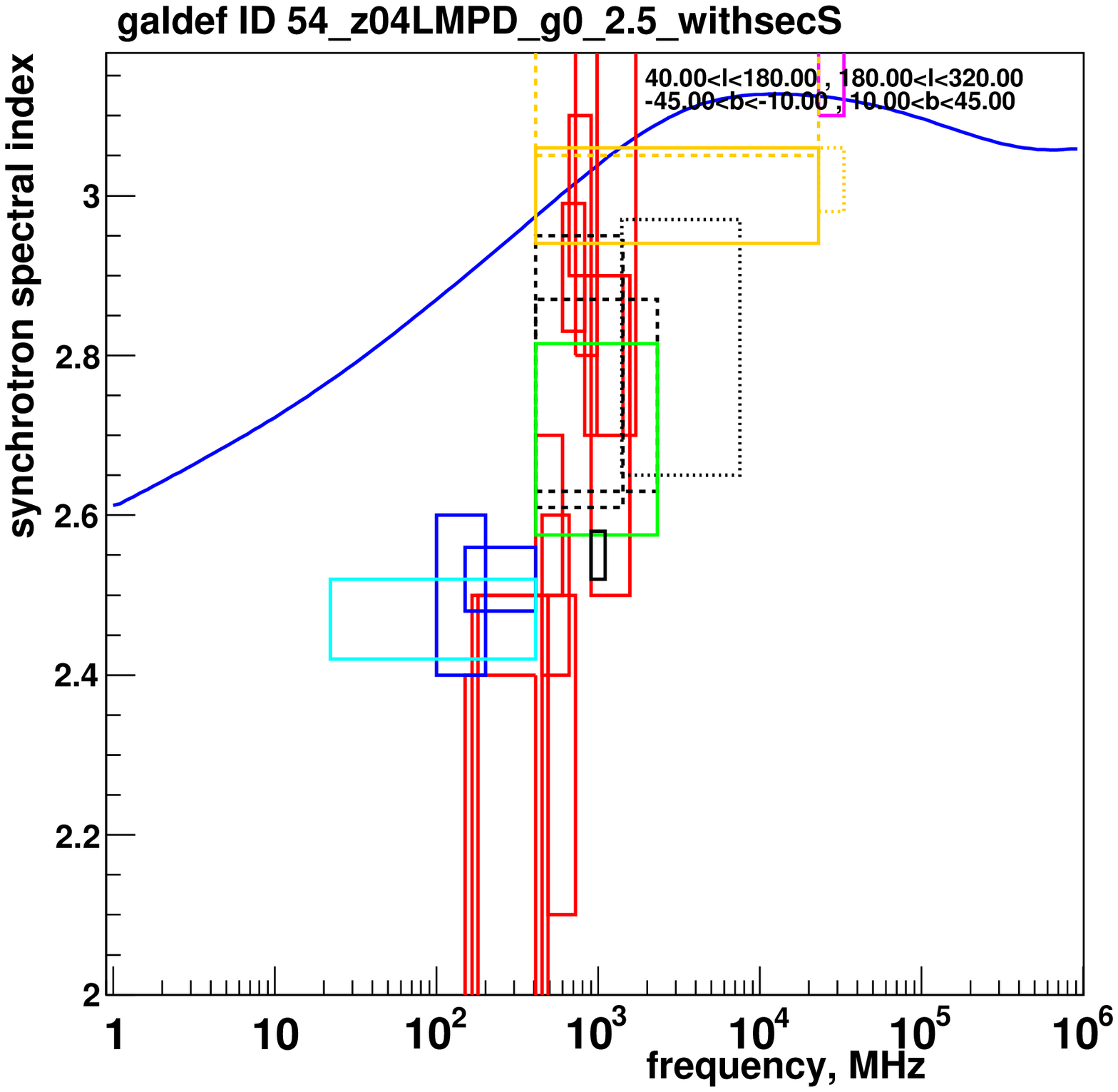}
\caption{ Synchrotron spectral index for pure diffusion model with low-energy electron injection index (left to right, top to bottom) 1.0, 1.3, 1.6, 1.8, 2.0, 2.5. Including secondary leptons. Experimental ranges are based on the references reviewed in Section 4.1, and are intended to be representative not exhaustive.   Data as in figure~\ref{PD_sync_indices_basic}.}
\label{PD_sync_indices_various}
\end{figure}

In Fig~\ref{PD_sync_spectra_cutoff}, the primary electron spectrum has been cut off below 4 GeV to illustrate the contribution from those energies; since removing these low-energy electrons eliminates most of the low-frequency synchrotron from primaries, it shows that low-frequencies (below 100 MHz)  are dominated by leptons with energies less than 4 GeV.
 Secondary leptons produce one third of the observed low-frequency intensity and hence make determination of the primary spectrum more difficult. Secondary leptons together with primaries above 4 GeV already account for 50\% of the low-frequency synchrotron.

\begin{figure}
\centering
\includegraphics[width=0.43\textwidth, angle=0]{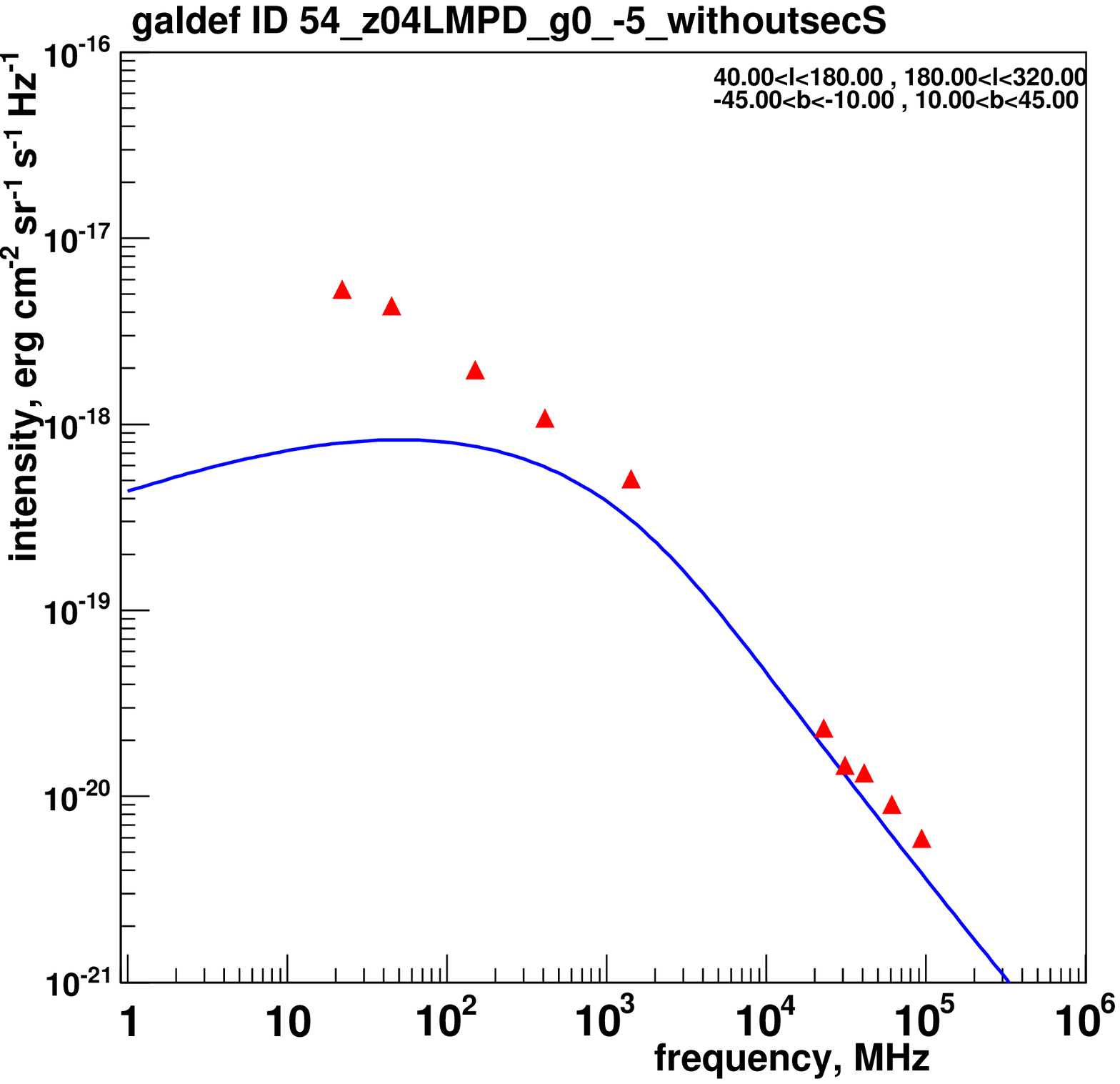}
\includegraphics[width=0.43\textwidth, angle=0]{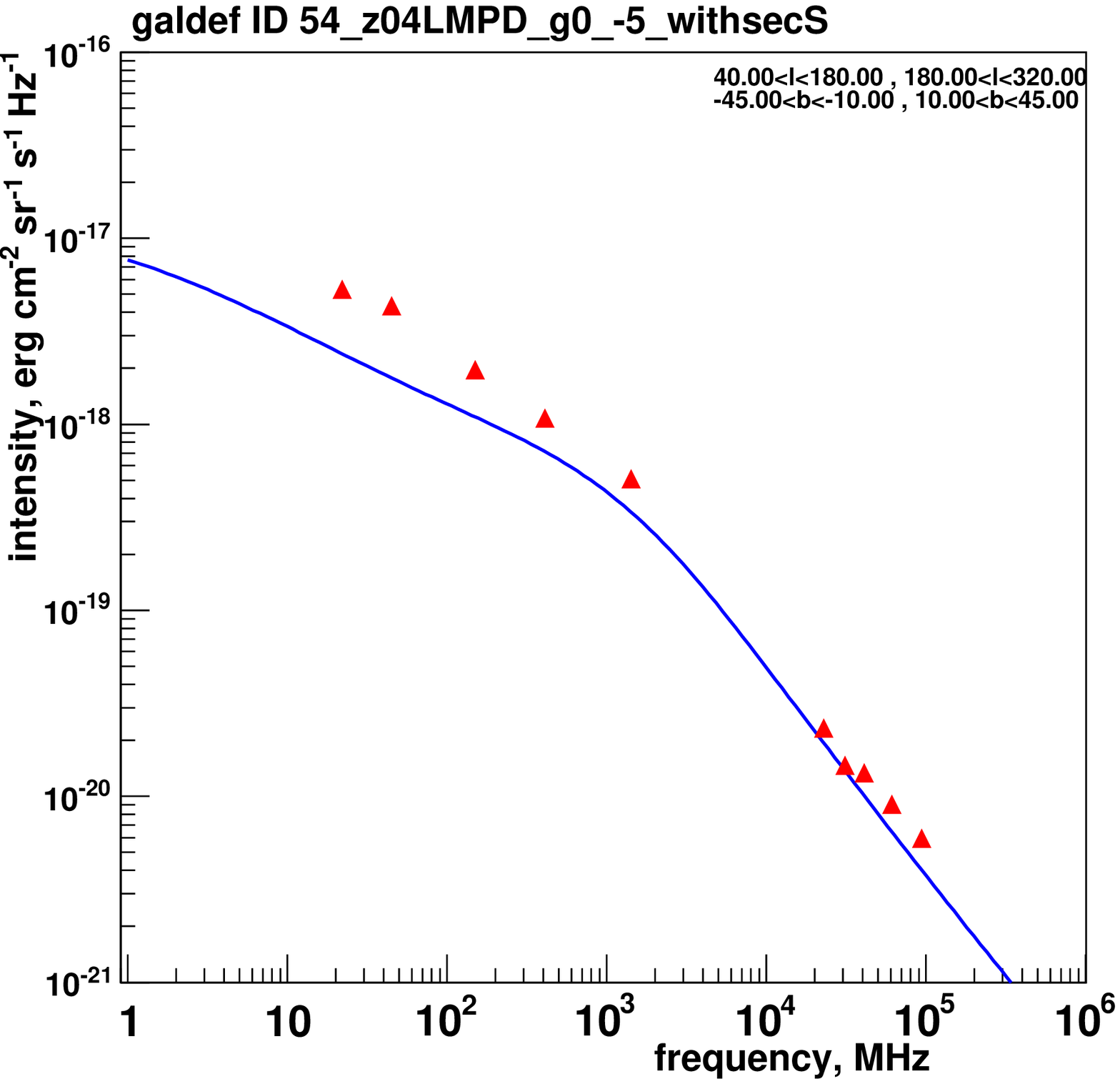}
\caption{Synchrotron spectra for pure diffusion model with sharp cutoff in primary electrons below 4 GeV; primary electrons only (upper),  primary and secondary leptons (bottom). The contribution from secondary leptons is shown in Fig 1.  Data as in figure~\ref{PD_sync_spectra_basic}. }
\label{PD_sync_spectra_cutoff}
\end{figure}

\subsection{Reacceleration model}

We now consider a reaccleration model, also with halo height 4 kpc.
 The complete set of GALPROP parameters are given in \citet{2041-8205-722-1-L58}, model z04LMS; the injection spectral index above 4 GeV has been reduced from 2.42 in that model to 2.3 to better fit the Fermi electron data above 20 GeV.
The range 7--20 GeV shows a slight steepening in Fermi-LAT and PAMELA data, but no attempt has been made to reproduce this in the reaccelation model since we want to test an existing published model; it has no effect on our conclusions.
 As in the case of the pure diffusion model a cutoff above 2 TeV has been introduced to fit the H.E.S.S. data, although this has no significance for synchrotron.
The lepton and synchrotron spectra for this model are shown in Fig~\ref{reacc_lepton_spectra}, Fig~\ref{reacc_sync_spectra} and the synchrotron indices in  Fig~\ref{reacc_sync_indices}.

It is clear that this particular reacceleration model is not consistent with the observed synchrotron spectrum, since the sum of primary and secondary leptons produces too high intensities at low frequencies, and the low-frequency spectral index is too large. It could be adjusted by making the low-energy injection index smaller, as for the pure diffusion model. However a large part of the excess comes from the secondary leptons which have a large peak due to reacceleration which makes them equal to primary electrons around 1 GeV, and  which cannot be adjusted very much in this model; this peak is not present in the pure diffusion model (see comparison for secondary leptons in Fig~\ref{PD_sync_spectra_basic}). Decreasing the B-field can improve the low-frequency fit but then the high-frequencies are under-predicted since the overall spectrum is steeper than the model predicts.
Only if the secondaries are removed does the synchrotron give a good fit, while the secondary production is certainly present. Arguing differently, we note that the secondaries already produce the low-energy synchrotron intensities, which would preclude the existence of electron primaries. Either way the model is problematic for synchrotron. This does not mean that reacceleration models are excluded by this study, but it does pose a challenge for future work on such models. 
 Reacceleration is surely present at some level on physical grounds, but probably less than currently adopted on the basis of B/C data.

 This is an example where solar modulation can be invoked to get agreement of a model with directly-measured CR data (increasing the interstellar spectrum and modulation appropriately), but not for synchrotron which probes the interstellar spectrum.

\begin{figure}
\includegraphics[width=0.40\textwidth, angle=0]{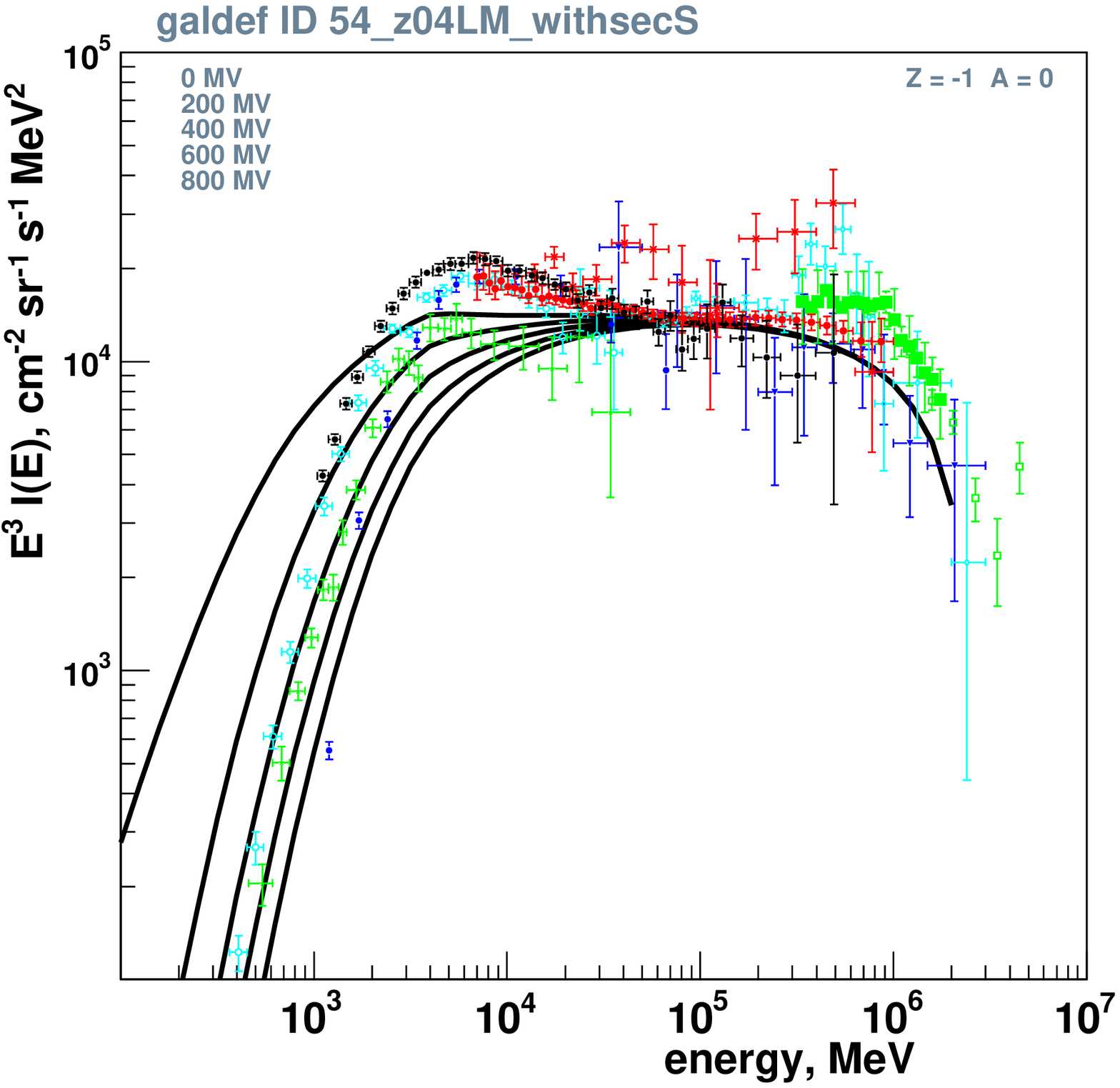}
\includegraphics[width=0.40\textwidth, angle=0]{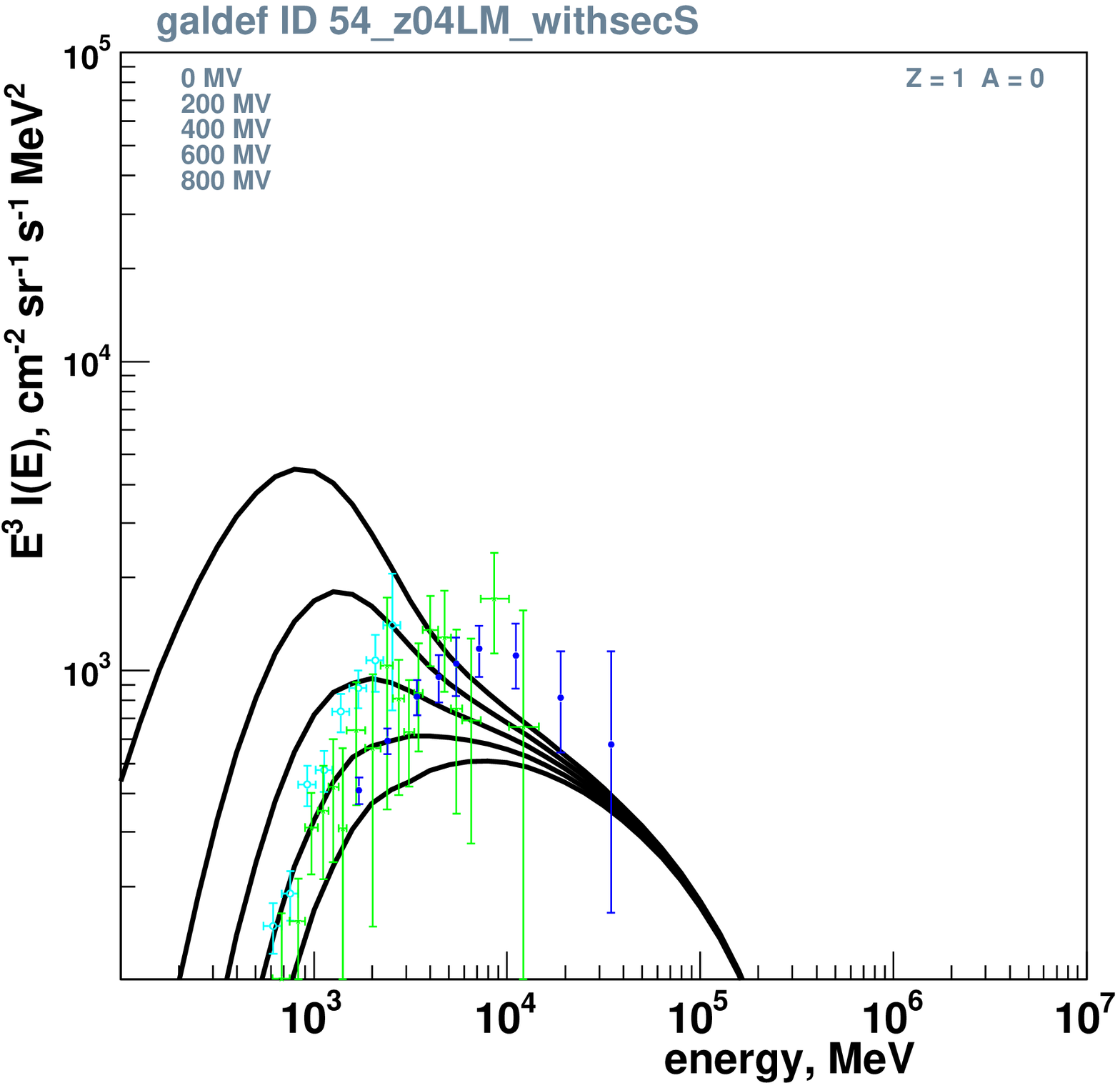}
\caption{Lepton spectra  for diffusive reacceration model with primary low-energy electron injection index 1.6.  Primary electrons  (upper), secondary positrons (lower).  Modulation $\Phi$=0,200,400,600,800 MV.  Data as in figure~\ref{PD_lepton_spectra_basic}. }
\label{reacc_lepton_spectra}
\end{figure}

\begin{figure}
\includegraphics[width=0.30\textwidth, angle=0]{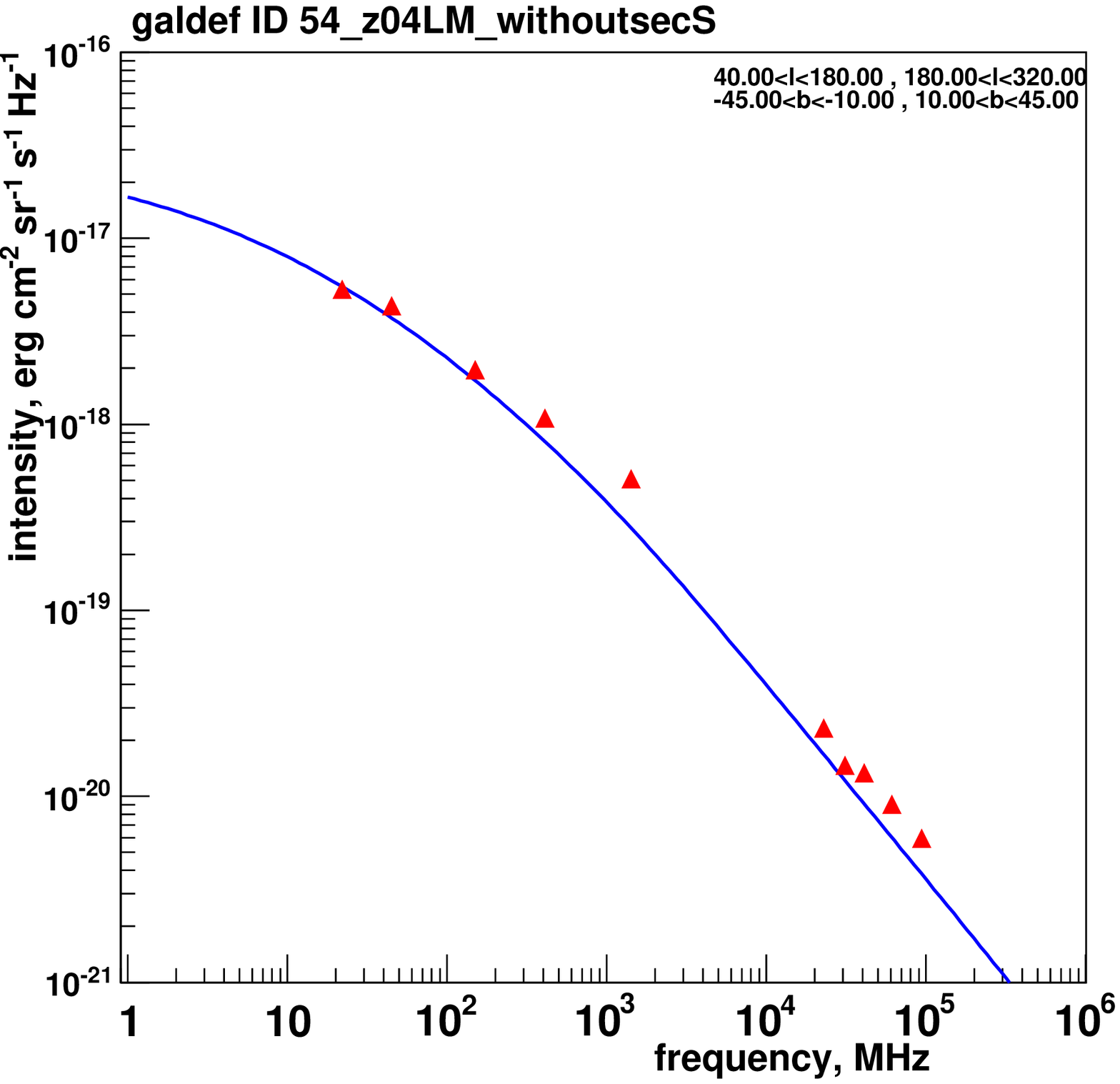}
\includegraphics[width=0.30\textwidth, angle=0]{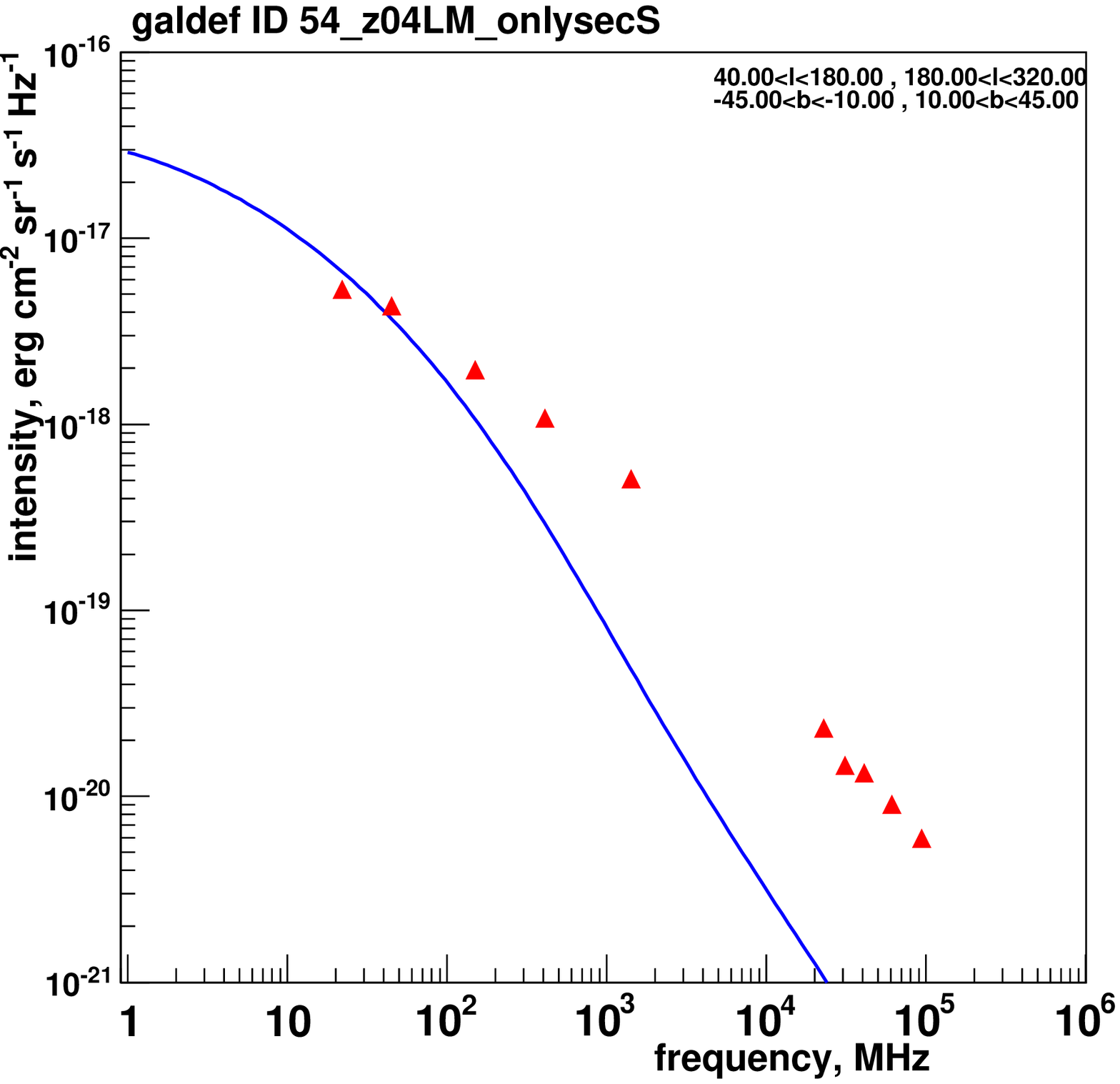}
\includegraphics[width=0.30\textwidth, angle=0]{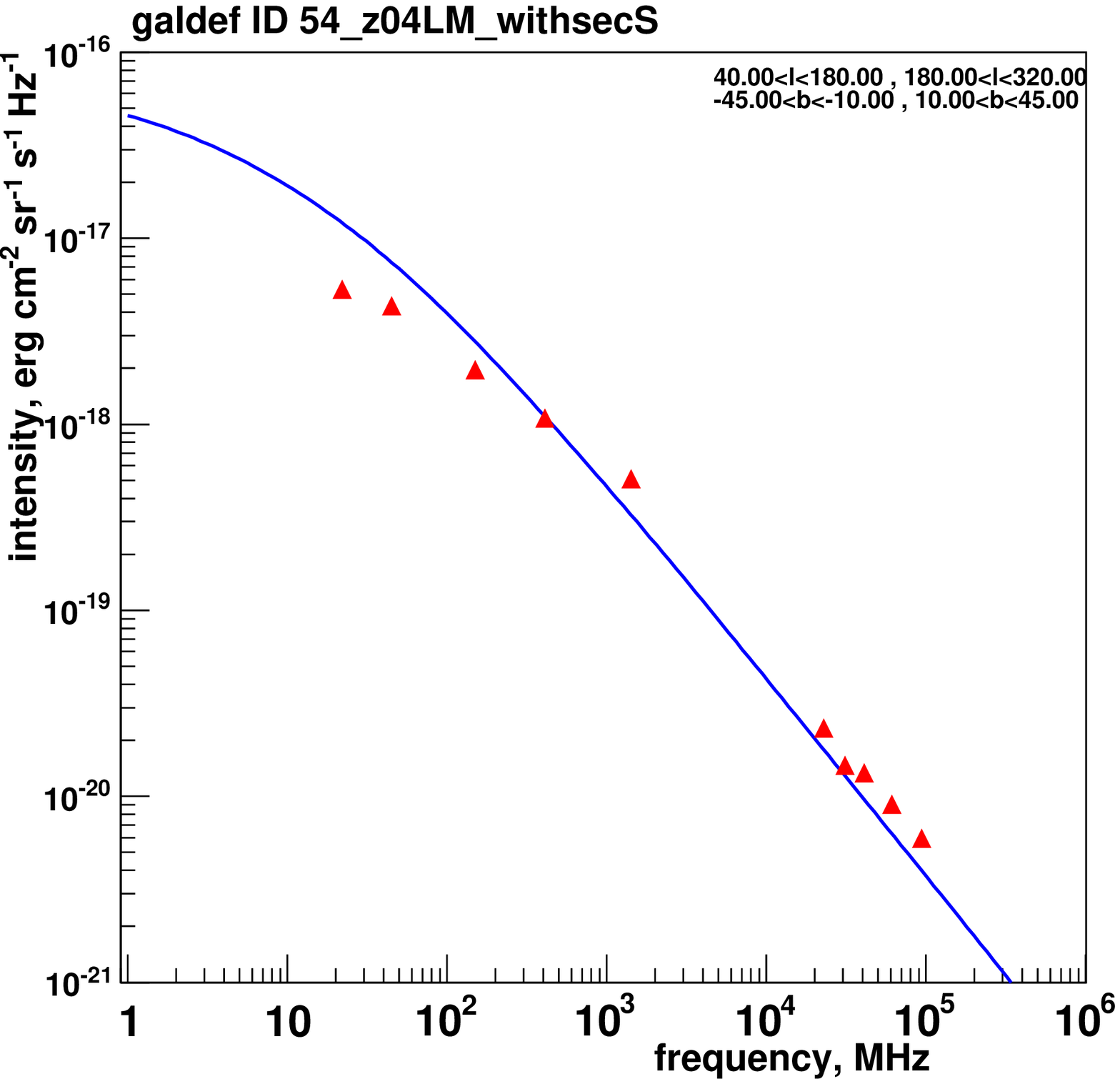}
\caption{ Synchrotron spectra for diffusive reacceration model with primary low-energy electron injection index 1.6. Synchrotron from primary electrons  (upper), secondary leptons (middle) and total (lower).  Data as in figure~\ref{PD_sync_spectra_basic}. }
\label{reacc_sync_spectra}
\end{figure}

\begin{figure}
\includegraphics[width=0.30\textwidth, angle=0]{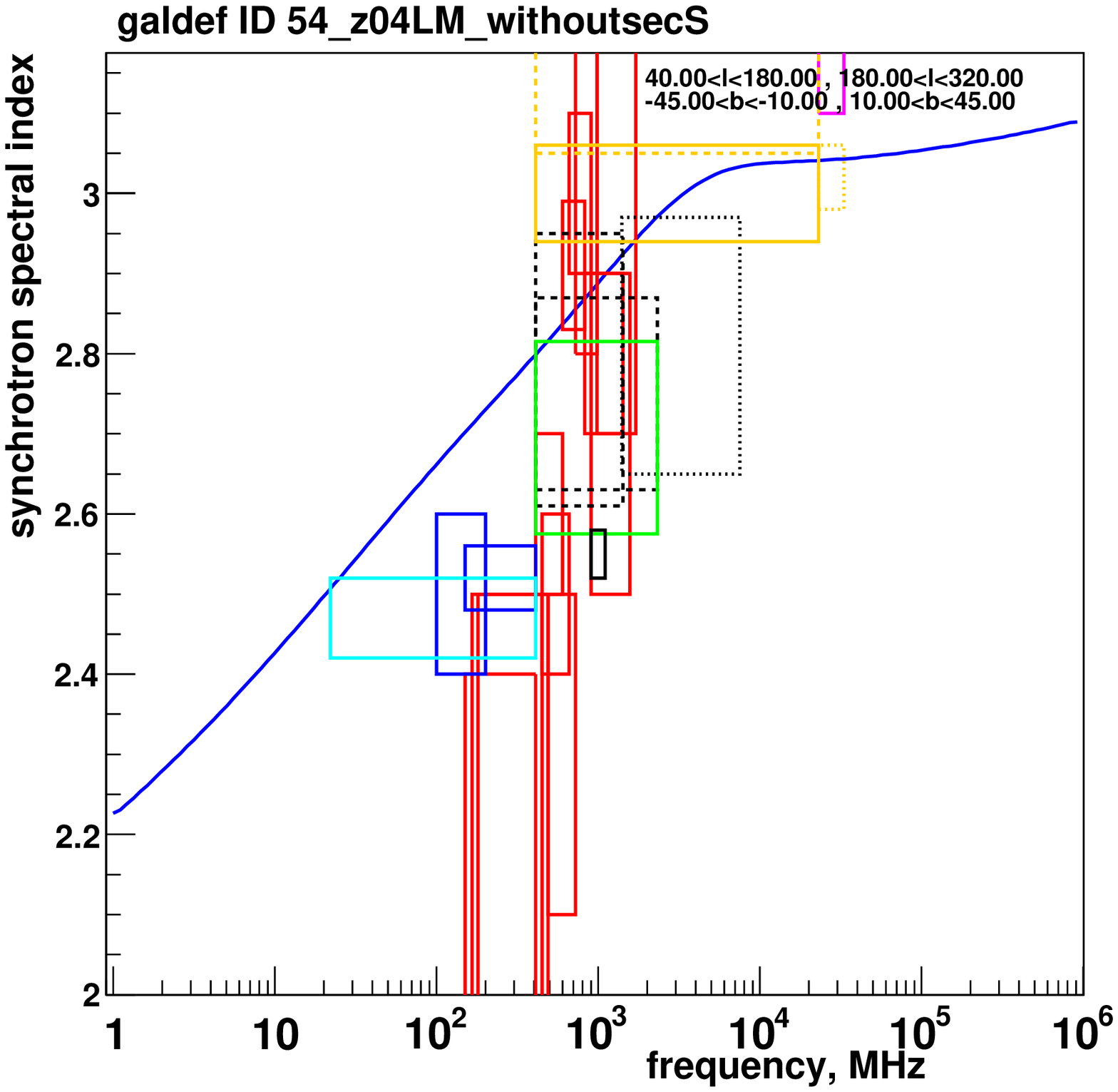}
\includegraphics[width=0.30\textwidth, angle=0]{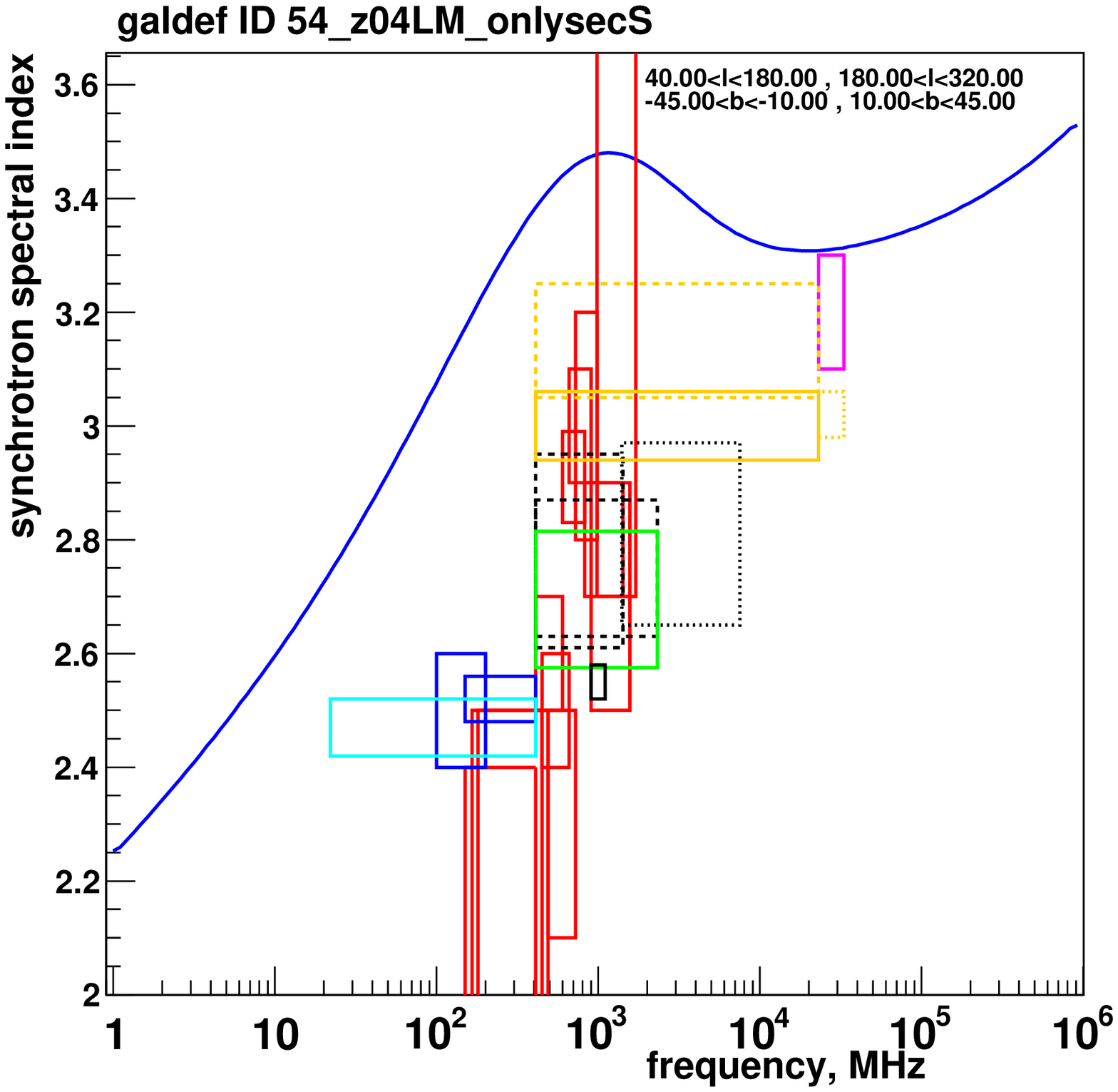}
\includegraphics[width=0.30\textwidth, angle=0]{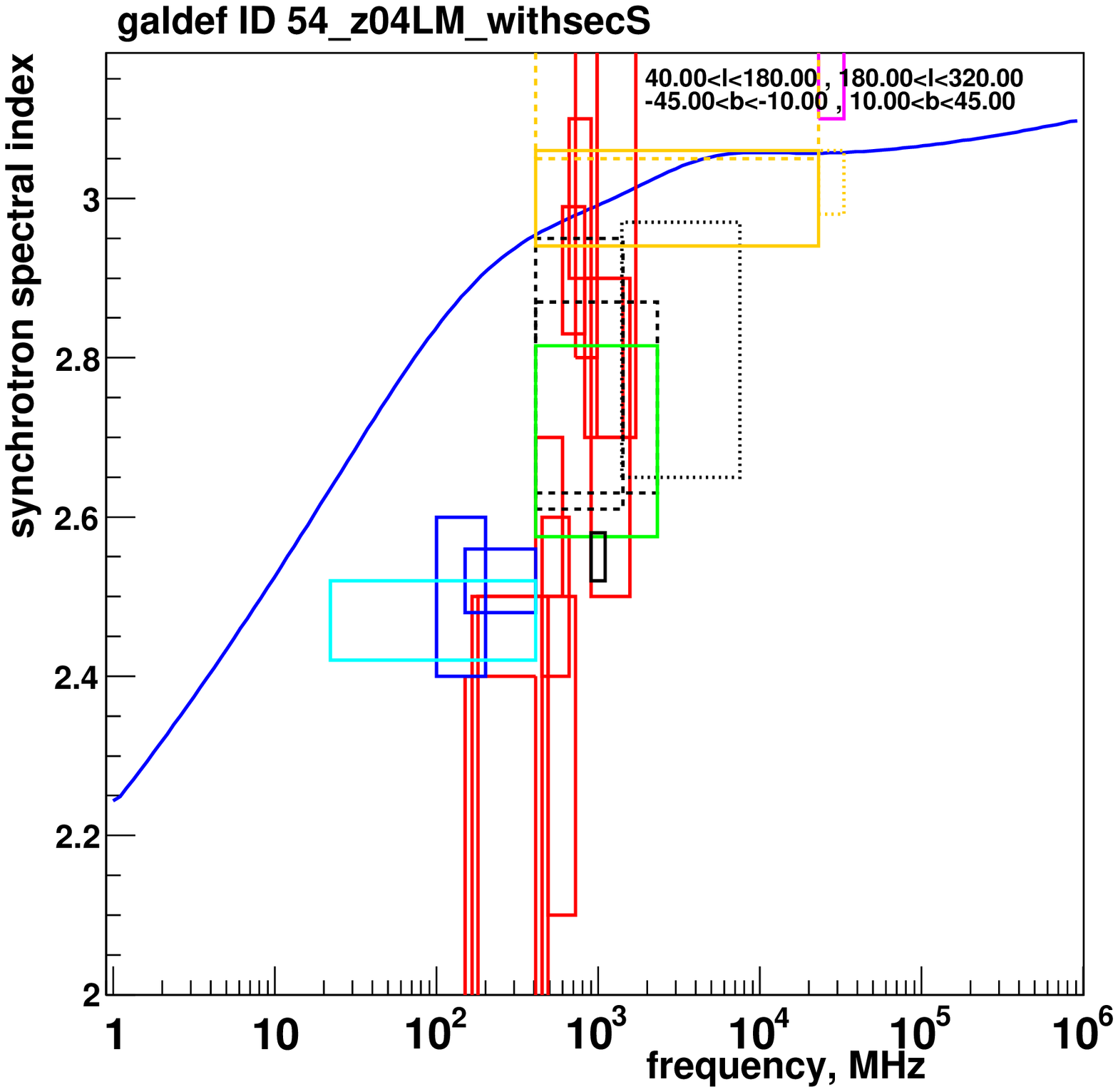}
\caption{ Synchrotron spectral index for diffusive reacceration model with primary low-energy electron injection index 1.6. Synchrotron from primary electrons  (upper), secondary leptons (middle) and total (lower).  Data as in figure~\ref{PD_sync_indices_basic}.  }
\label{reacc_sync_indices}
\end{figure}


\subsection{Low-energy spectral index from propagation ?}

 The low-energy injection indices deduced for the standard propagation models are unexpected (see Discussion),
so it is valid to ask whether they could be produced by propagation using a more `normal' injection spectrum with index 2.
One way to do this is to reduce the energy losses by making the propagation region smaller - which means reducing the halo size.
Then the steepening by propagation is reduced. This is at the expense of no more fitting B/C, $^{10}$Be/$^9$Be and local electron measurements,
but it is worthwhile illustrating this explicitly. Fig~\ref{PD_1kpc} shows a model with halo height 1 kpc, but otherwise the same as the previous pure diffusion model (with its low-energy injection index 2 and halo height 4 kpc).
(The diffusion coefficient has not been changed to fit B/C since this would increase the propagation time and the energy losses would be unchanged from the larger halo case, and the synchrotron spectrum would be the same as before.)    To get the correct synchrotron intensity, the electron flux has to be increased by 3 compared to that observed locally to compensate the smaller integration length (increasing B would just lead back to the same losses again). With this unnatural scenario we can indeed reproduce the observed synchrotron spectrum  (Fig~\ref{PD_1kpc}). However since several other constraints are thereby violated 
 (B/C : model too low,  $^{10}$Be/$^9$Be: model too high
\footnote{ B/C $\propto z_h/D(E)$, $^{10}$Be/$^9$Be $\propto \sqrt{ D(E)}/z_h$.   }
, local electron spectrum : model too high)
 this shows the difficulty of constructing such a model consistently. There are of course other possibilities to approach this issue, but we restrict ourselves to this example here.

\begin{figure}
\includegraphics[width=0.45\textwidth, angle=0]{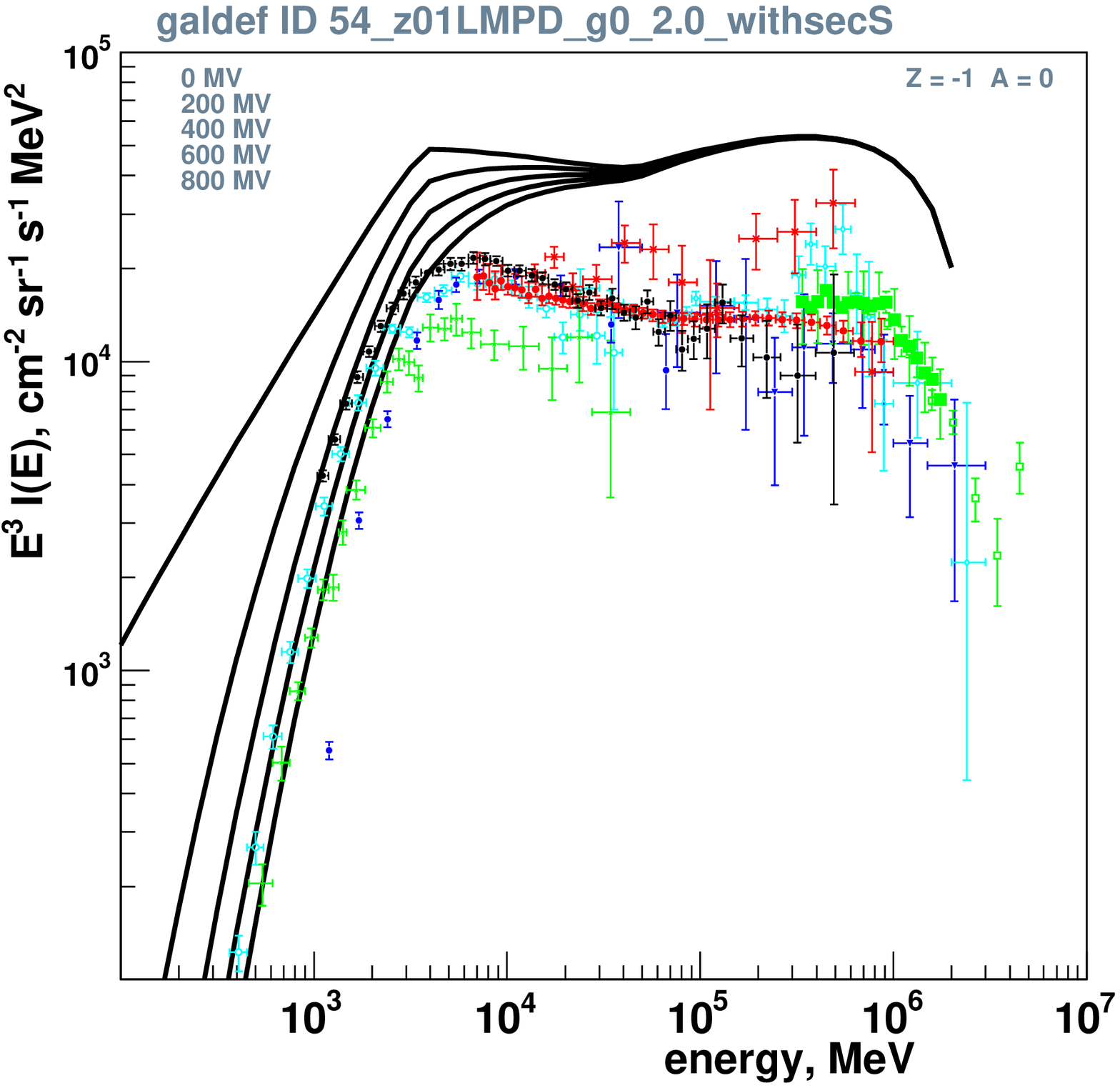}
\includegraphics[width=0.45\textwidth, angle=0]{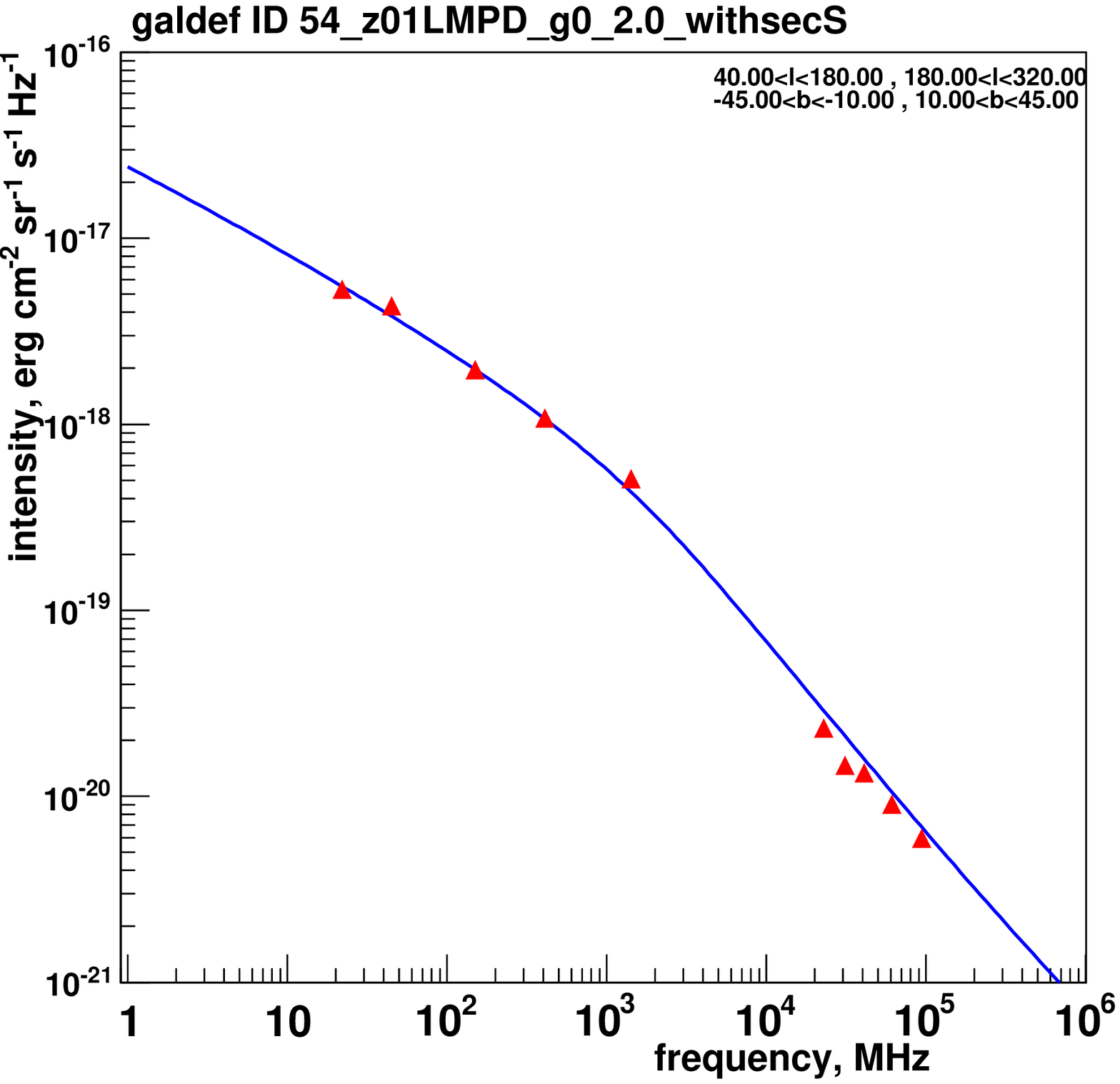}
\includegraphics[width=0.45\textwidth, angle=0]{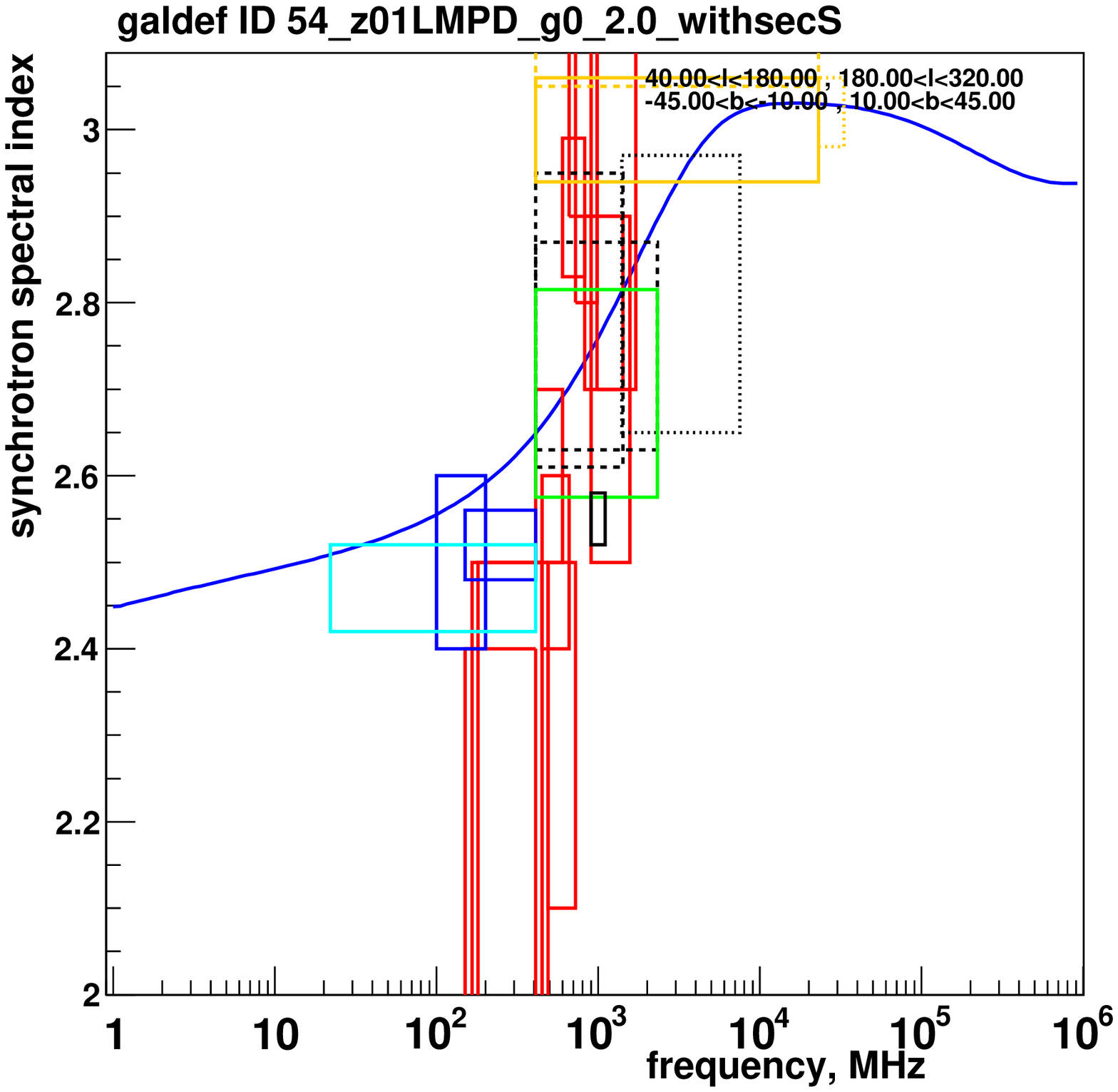}
\caption{Top: Electron spectrum for pure diffusion model with halo height 1 kpc, low-energy electron injection index 2.0. Modulation $\Phi$=0,200,400,600,800 MV. Data as in figure~\ref{PD_lepton_spectra_basic}. Centre and bottom : Corresponding synchrotron spectrum and spectral index, plotted as in  figure~\ref{PD_sync_spectra_various}, with which this figure should be compared. }
\label{PD_1kpc}
\end{figure}

 Another alternative to obtain the electron spectrum by propagation is to invoke an upturn in D(E) at low energies, instead of the constant used normally. 
 Fig~\ref{PD_Dmod} shows a model with   $D(E)\propto E^{-0.5}$ for E $<$ 4 GeV,  but otherwise the same as the pure diffusion model with low-energy injection index 2 and halo height 4 kpc.
It reproduces the synchrotron data, and does not require the high electron spectrum of the model with small halo height.
 However it will under-predict the B/C data at low energies - but this depends strongly on solar modulation and hence is not so critical.
 It is a possibly a more plausible scenario than the previous one since it violates less constraints.
The required D(E) is in fact similar to that in  the wave-damping model of
\cite{2006ApJ...642..902P} 
and hence has also a plausible physical basis.

\begin{figure}
\includegraphics[width=0.45\textwidth, angle=0]{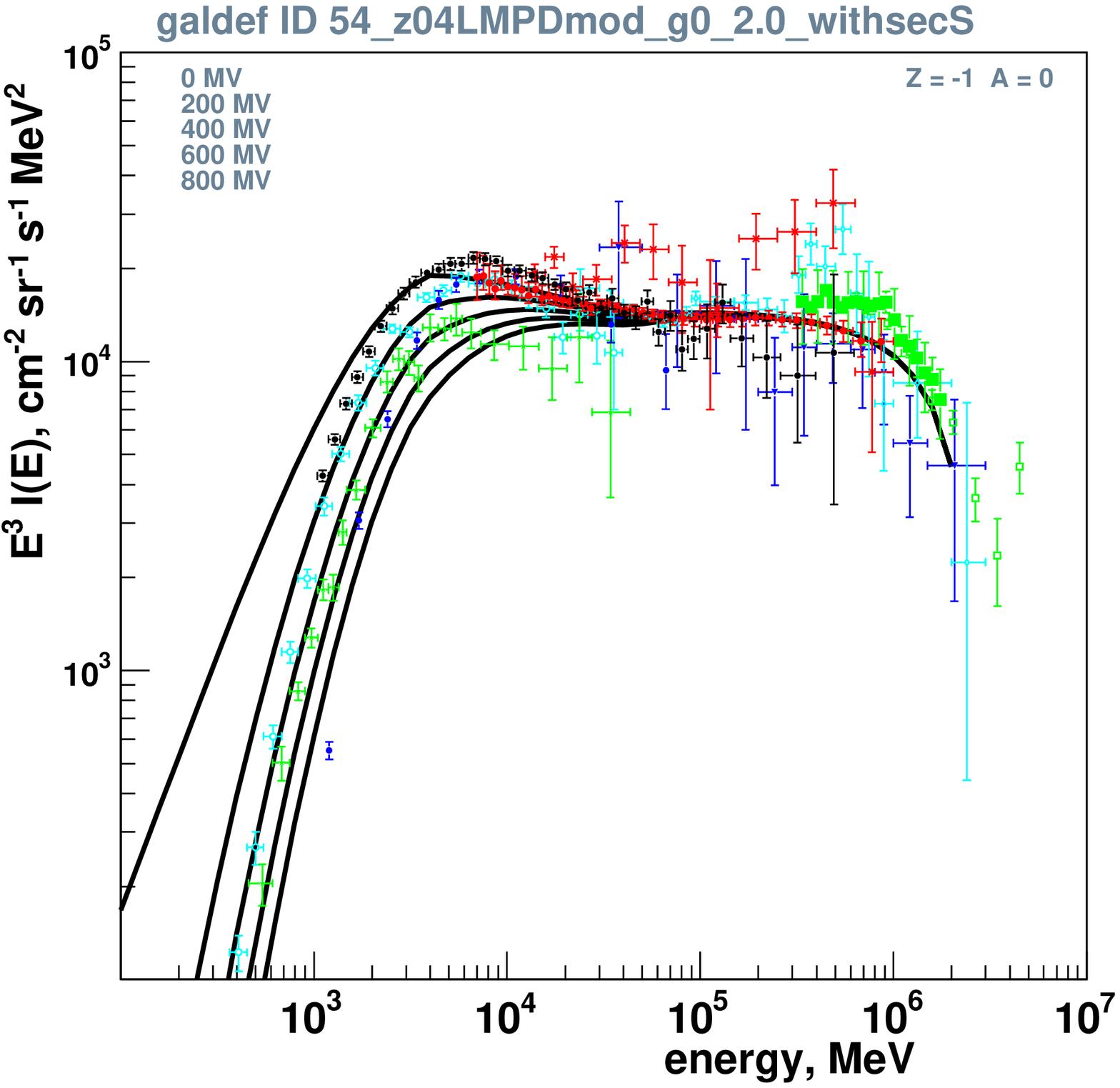}
\includegraphics[width=0.45\textwidth, angle=0]{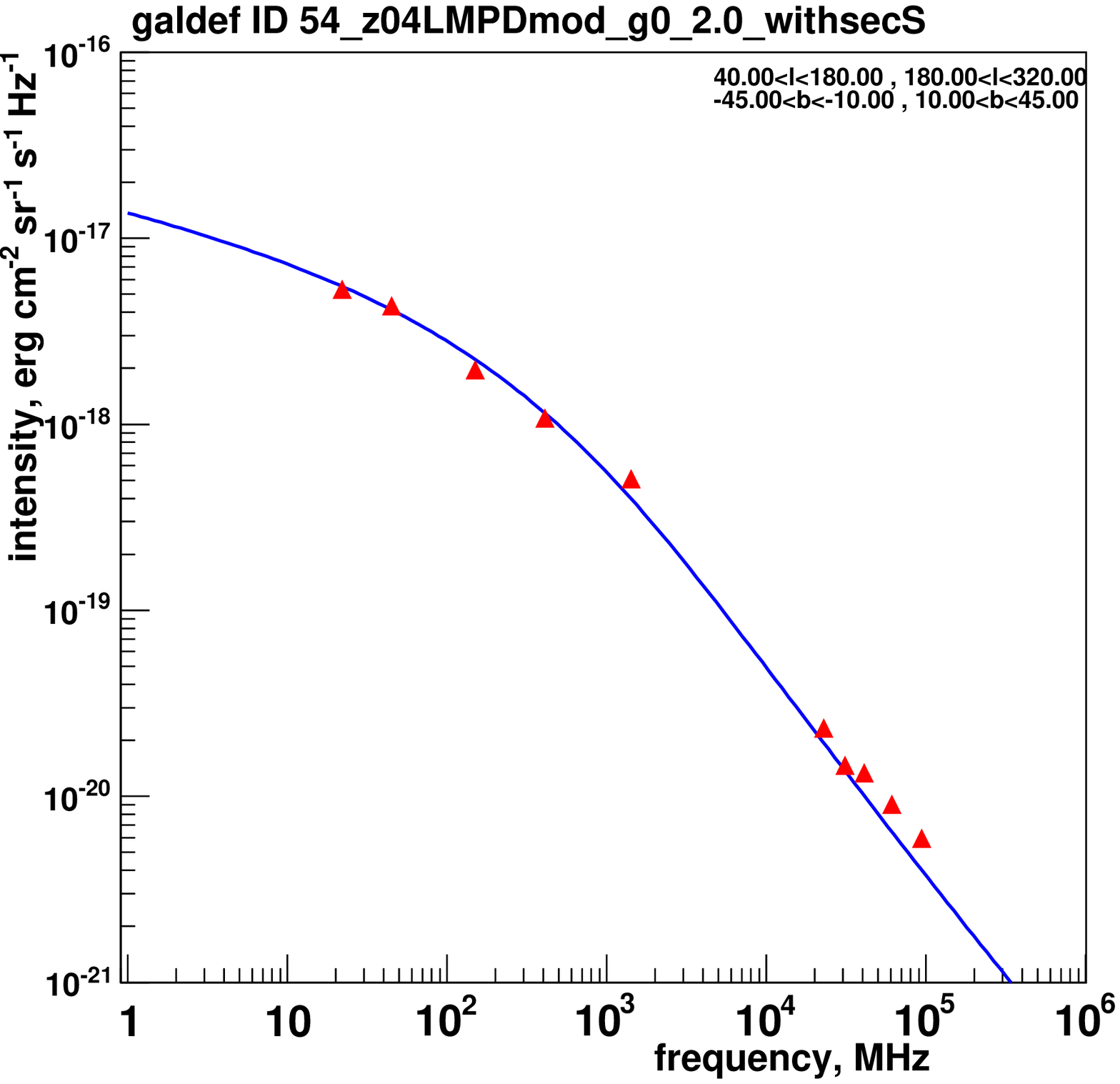}
\includegraphics[width=0.45\textwidth, angle=0]{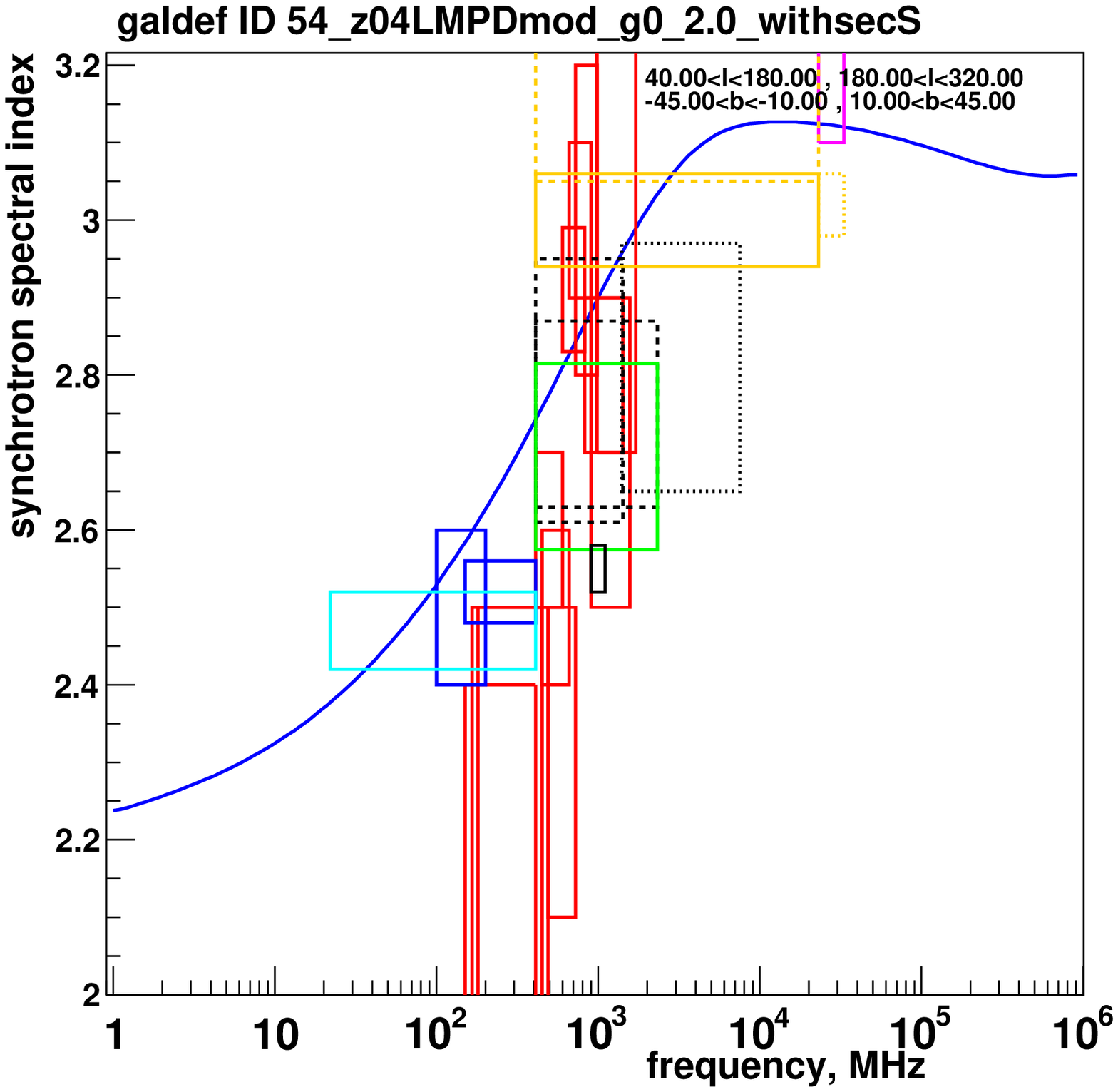}
\caption{Top: Electron spectrum for pure diffusion model with low-energy $D(E)\propto E^{-0.5}$, halo height 4 kpc, low-energy electron injection index 2.0. Modulation $\Phi$=0,200,400,600,800 MV. Data as in figure~\ref{PD_lepton_spectra_basic}. Centre and bottom : Corresponding synchrotron spectrum and spectral index, plotted as in  figure~\ref{PD_sync_spectra_various}, with which this figure should be compared. }
\label{PD_Dmod}
\end{figure}

 Other variations on the propagation could affect the electron spectrum, for example spatial variations in the diffusion coefficient, anisotropic diffusion and convection, which have not been considered here. They will not however affect our general conclusion of the need for a significant break in the electron injection spectrum. This is because any model must be constrained by the CR nuclei secondary/primary data, and the resulting modified propagation parameters will finally lead to a path-length distribution similar to our basic model, and hence similar electron energy losses.

  Although we might expect a dependence of D(E) on the B-field, in these models the variation of total B-field is very small (see section 2.3), so that including such a dependence would have no sigificant effect. However in future such a dependence on B and its topology  could be included to make the models more physically realistic.

 High-energy electrons are expected to have spatial variations due to their rapid energy losses combined with the stochastic nature of the sources in space and time. A study of this effect \citep{2001ICRC....5.1964S} shows that significant variations start above 10 GeV, and become large only above 100 GeV, and hence will not affect the low-energy synchrotron from lower energies which is the main focus of this work. In any case the synchrotron spectra we use here are integrations over large sky areas and long lines-of-sight, which will smooth out any variations and hence not affect our conclusions.

\section{Discussion}

Independent of the propagation model, 
the primary  electron spectrum must turn over below a few GeV, with an ambient index around 2. In the pure diffusion model this  implies an injection index 1.3-1.6.
It cannot cutoff completely since secondary leptons only contribute about one third of the low-frequency synchrotron, and electrons below 1 GeV are anyway directly observed by spacecraft in the heliosphere.

While the synchrotron spectra may be subject to zero-level and scale errors, and the effects of absorption and free-free emission may affect the comparison with models, the determination of the spectral index has been performed in a robust way by many authors and this gives a rather tight constraint on the ambient electron spectrum. 
The most recent determination of the spectral index 45-408 MHz \citep{2011A&A...525A.138G} 
 gives 2.5-2.6, implying an ambient electron index\footnote{The synchrotron spectral index for an ambient electron spectrum with power-law index $p$ is approximately $\beta$=2+($p$-1)/2.} of  2.0-2.2 for electrons below a few GeV.
 This completely excludes a continuation of the ambient electron index 3.0-3.2 measured by Fermi-LAT $>$7 GeV) to energies below a few GeV.
 Since in this range the diffusion coefficient is constant in the pure diffusion model, only energy losses steepen the spectrum slightly since escape dominates. This therefore implies an electron injection index of $<2$, consistent with the detailed analysis presented here. While the latter is a very simplified argument it supports the conclusions of the detailed calculations in a  robust model-independent way.
At the same time the synchrotron index $>$ 1 GHz has been found by many authors to be near 3, fully consistent with the measured ambient electron spectrum above a few GeV (ambient index 3.0-3.2 giving synchrotron index 3.0-3.1), and an injection index 2.2 steepened by 0.5 from synchrotron/IC losses and by 0.5 due to D(E).

 To be more precise on this point, we can use
an analytical  approximation to the propagated electron spectrum for a plane parallel source distribution with diffusion (no reacceleration) and energy losses given e.g. by
\cite{1974Ap&SS..29..305B} 
and
\cite{2010A&A...524A..51D}. 
The spectrum steepens by $(\delta + \alpha - 1)/2$ where dE/dt $\propto E^{\alpha}$ and D(E)$\propto E^\delta$.
For synchrotron losses $\alpha=2$, so at high energies where $\delta=0.5$ the steepening is 0.75, in accordance with our injection spectrum 2.2 and ambient spectrum 3,
and at low energies  $\delta=0$ the steepening is 0.50, consistent with our injection index 1.3 - 1.6 and ambient spectrum 2. 
In reality the relations are more complex due to spatial dependence etc., but the analytical form reproduces the general behaviour for synchrotron/IC-type losses, and shows it does not depend critically on the detailed GALPROP modelling.

Again we note that the determination of the ambient electron spectrum is independent of the way in which this spectrum was produced via injection and propagation. We use a particular type of parameterized model to generate physically plausible spectra, but other models would lead to the same ambient spectrum since the final criterion is consistency with synchrotron and direct measurements.

The low-energy interstellar spectrum has consequences for solar modulation: it must have a smaller effect than predicted by  the force-field approximation for $\Phi$=600--700 MV generally used for such solar-mininum data  (see Section 5); using our synchrotron-based spectrum,   $\Phi\leq$200 MV gives a better fit to  data $<$2 GeV. The new PAMELA data reinforces this conclusion. Since the force-field approximation is anyway known to be unreliable, and the values used are often based on an assumed interstellar spectrum,  we do not pursue this further, but simply propose that the interstellar spectrum determined in this paper be used in future physical modelling of modulation.
 From our analysis we just suggest that the low-energy falloff in the directly measured electrons,  normally attributed mainly  to modulation,  may instead reflect more the interstellar spectrum.
We note that although synchrotron probes  CR leptons  but not nuclei, the improved understanding of modulation would be relevant to all species.

 It is intentionally beyond the scope of this paper to speculate on the origin of the electron injection spectrum; we simply present it as an observational result posing a challenge for cosmic-ray source models. However some obvious remarks are in order. The low-energy injection spectrum is less than found in SNR;
the distribution of SNR radio spectral indices 
\citet{2010A&A...524A..51D} 
 is very broad - index 2.2 to 2.8, with mean 2.48 giving an electron index 1.4 to 2.6, with mean 1.96. They quote an SNR electron index 2.0$\pm$0.3.
However this is for electrons inside the SNRs: the escaping flux may be harder if higher-energy electrons escape
more easily as might be expected. At high  energies our derived injection index 2.2 suggests
a steepening relative to the spectrum inside the remnants, but at lower energies this may not hold.
A recent study of escape of electrons from SNR is given by
 Ohira et al. 2011 (arXiv:1106.1810),
and \cite{2010A&A...513A..17O} ,
which includes among many effects an escape time decreasing with energy, and hence a hardening of the escape spectrum.
  Our results will constrain such models, which predict a complex escape spectrum.

Other sources of electrons may also contribute, for example pulsar wind nebula (PWN) have radio indices 2-2.3,
\citep{2011SSRv..tmp...46R} 
 giving electron index 1-1.6, similar to our low-energy injection index. While PWN clearly
cannot produce the high-energy spectrum, it is possible that they play a role at low energies.
Pulsars produce electrons with index 1.5 - 1.9 (with a cutoff at several GeV)
 \citep{2011NIMPA.630...48G} 
again similar to our low-energy index.
The attempt to construct  a model with various source types to reproduce the data is beyond the scope of the present work,
and in fact it is not easy to imagine how hard spectrum low-energy sources plus steeper spectrum high-energy sources could be combined
to produce the observed composite spectrum.

 Finally we compare the electron injection spectrum with that of nuclei, in particular protons. In the GALPROP models used here, the 
nuclei injection spectrum has index,  above/below a break energy of 9 GeV/nucleon, of  1.8/2.25 for the plain diffusion model, and 1.98/2.42 for the reacceleration model.
 These values have been chosen to agree with CR data for the given propagation parameters derived from B/C etc. Thus the low-energy nuclei index is slighty larger
than what we have deduced for electrons (1.3-1.6 for the plain diffusion model, 1.6 for the reacceleration model).
 However the nuclei are strongly affected by solar modulation at low energies, and
unlike the electrons there is no equivalent of the synchrotron tracer for the interstellar spectrum 
(pion-decay gamma rays may be constraining but are sensitive mainly to protons above a few GeV).
 So a nuclei injection spectrum equal to that of electrons would be acceptable,
if indeed that should be predicted by a theory of CR acceleration, but is not required by the data.

\section{Conclusions}

Our main conclusion is that the  interstellar CR electron spectrum must turn over rather sharply below a few GeV.
This result is independent of how the spectrum is formed by injection and propagation.
The low-energy falloff in the directly measured electrons,  normally attributed just  to modulation,  may instead reflect mainly the interstellar spectrum.
The (model-dependent) injection index implied for the primary electrons is 1.3-1.6 below a few GeV, and 2.1-2.3 at higher energies.
The standard reacceleration model is not consistent with the observed synchrotron spectrum, since the total from primary and secondary leptons exceeds the measured synchrotron at low frequencies. While not excluding reacceleration models, it does pose a challenge to be addressed.

We show that it is still possible to obtain the ambient electron spectrum by propagation even for a more conventional injection index of 2, but at the expense of violating other constraints. A low-energy upturn in the diffusion coefficient is the most promising model of this kind.

Therefore combining synchrotron data with direct measurements of CR provide unique and essential  constraints on the interstellar electron spectrum.
These results have implications for interstellar gamma rays especially at low energies  \citep{2008ApJ...682..400P}, and also for high energies (see \citet{2011arXiv1101.1381S} for a recent review).
 Exploiting the complementary information on  cosmic rays and synchrotron gives new constraints and has implications for gamma rays. This connection is especially relevant now in view of the ongoing PLANCK and Fermi missions, and in future new radio astronomy instruments like LOFAR and direct measurements by AMS-02.

  There are of course limits to what conclusions based on a model like GALPROP can achieve since the full complexity of the Galaxy and physical processes can never be reproduced. A key is the combination of constraints from many different data types (`multi-messenger'), but still it is hard to break the degeneracy between the source injection spectrum and propagation even with secondary/primary ratios etc. In future as CR sources are better understood via radio, X-ray  and  gamma-ray observations we should be able to make more basic progress on this topic.

\begin{acknowledgements}

We acknowledge useful comments  from I. Moskalenko, T. Porter, G. J\'ohannesson and A. Vladimirov. We thank the referee for critical comments which helped to improve this paper.

\end{acknowledgements}
\bibliographystyle{aa}
\bibliography{synch,luminosity,strong}

109, 471 



\begin{appendix}
\section{Radio spectral index data.}

Starting with lower frequencies,
 \citet{1999A&AS..137....7R} 
 used the 22 MHz and 408 MHz surveys to derive average high latitude $\beta=2.47$, with a variation of about 0.05.
\citet{2008AJ....136..641R}  
measured over a continuous band 100-200 MHz, $\beta=2.5\pm0.1$ ; combining with other surveys 150-408 MHz $\beta=2.52\pm0.04$ at high latitudes. This paper contains a useful summary table of $\beta$ back to 1962.
\citet{2011A&A...525A.138G} 
use the recent 45 MHz all-sky map together with the 408 MHz Haslam map to derive   $\beta=2.5-2.6$ over most of the sky for these frequencies.

At higher frequencies,
\citet{1988A&A...196..211R},  
separating thermal and nonthermal  components for 408-1420 MHz found a non-thermal index $\beta=2.85-3.1$.
\citet{2004mim..proc...63R}  
give spectral indices for 45-408-1420-22800 MHz, and give a table of zero-level corrections. Spectral index maps.  $\beta(408-1420 MHz ) = 2.6-2.7$ away from plane and loops.
\citet{2002A&A...387...82G} 
 find 408-1420 MHz $\beta=2.78\pm0.17$ and  408-2326 MHz $\beta=2.75\pm0.12$.
\citet{2003A&A...410..847P}  
  from a full-sky analysis of 408, 1420 and 2326 MHz survey find  $\beta=2.695$ with a dispersion of 0.12.
\citet{1998ApJ...505..473P} 
using  radiometers at a high-altitude site at 1400-7500 MHz: $\beta=2.81\pm0.16$.
\citet{2008ApJ...688...12Z} 
 report TRIS  absolute measurements at 0.6, 0.82, 2.5 GHz , drift scans at $\delta-42^o$ (this is paper I, II=\citet{2008ApJ...688...24G}, III=\citet{2008ApJ...688...32T}).
\citet{2008ApJ...682..223G} 
derive the extragalactic source contribution to the background from 151 to 8440 MHz.
\citet{2008ApJ...688...32T} 
 using TRIS III find  a synchrotron halo with $\beta$=2.9-3.1 600-820 MHz. This paper discusses zero errors in other surveys (their table 9: 150, 408, 820, 1420 MHz surveys). They find that  $\beta$ increases from 2.2 to 2.8 from 150 MHz to 1420 MHz. They give a spectrum in two directions (9h,42$^o$, 10h,42$^o$) corrected for zero level (their fig 7) which can serve as a standard: it shows large steepening from 150 to 1420 MHz). They give the variation of $\beta$(600-820 MHz)  along $\delta=42^o$ : it is mainly in the range 2.8 to 3.2(their fig 5). This paper contains an extensive discussion of spectral indices past and present, and its relation to electron spectrum.
Tgal=T-Toff=T-(Tex+Tcmb)+Tzero.
From their Table 9, the correction to the 408 MHz Haslam survey is Tzero=+3.9; EX=2.65 CMB=2.82 gives Toff=1.57
 compared to \citet{1988A&AS...74....7R} 
 (their table 7): Tzero=+2.1,Toff=$3.7\pm0.85$.


\citet{2011ApJ...734....4K}  
 (ARCADE2) at 3, 8 , 10 GHz :  $\beta_{sync}=2.55\pm0.03$  using a 408 MHz template, but they use cosec(b) and CII correlations.
Poles/coldest regions  $\beta_{sync}=2.57\pm0.03$
These values are significantly lower than those normally found (see above: 2.7 - 3.1).
\footnote{ In a related ARCADE2 study,
\citet{2011ApJ...734....5F} 
claim a  3-90 GHz extragalactic background  about a factor 6 higher than expected from radio sources at 1 GHz. Excess claimed to have   $\beta$=2.60 from 22 MHz to 10 GHz. To model the Galactic emission they use a cosec(b) analysis, which would not be sensitive to a large halo (see their discussion, correlation with CII).
This was followed up in
\cite{2011ApJ...734....6S}. 
who attribute the excess to underestimated Galactic emission or unaccounted radio sources, or some combination of both.
\citet{2011arXiv1102.0814V} 
use radio source counts to show that the claimed extragalactic background is hard to explain as the sum of sources to the currently observed flux limits, and that a extra population at lower fluxes would be required.
See also \citet{2010MNRAS.409.1172S}, 
 who also relate the radio to the extragalactic X- and gamma-ray backgrounds.
}

Turning to the highest frequencies,
\citet{2007ApJS..170..288H} 
using WMAP 3-year data found $\beta=3.15-3.5$ for 23 - 61 GHz.
\citet{2009ApJ...701.1804D}  
 used WMAP 5-year polarized maps to derive synchrotron $\beta=3.02\pm0.04$ (at WMAP frequencies). They show skymaps of  $\beta$, and find no latitude dependence unlike \citet{2007ApJ...665..355K} 
 (WMAP 3-year data) who found an increase from 3.05 to 3.25 from the Galactic plane to the poles.
\citet{2009ApJS..180..265G} 
 used WMAP 5-year data to derive  synchrotron $\beta=3.15\pm0.10$ (roughly from their fig 16), including polarized-only analysis. The latitude profile shows a lower index for $|b|<10^o$: 2.8,  (unlike \citet{2009ApJ...701.1804D} who found no latitude variation on the same 5-year data, but like \citet{2007ApJ...665..355K} who found similar variation slightly different values) on the 3-year WMAP data).
\citet{2008A&A...490.1093M}  
 give maps of 23 GHz intensity and $\beta$ 408 MHz-23 GHz using WMAP polarized emission maps, especially useful for comparing with synchrotron models since it selects out the (polarized) synchrotron emission in a model-independent way.
\citet{2008A&A...484..733T} 
 give spectral index distributions using polarized intensity for the 1.435 GHz survey and 22 GHz WMAP. It peaks at $\beta=2.7-3.0$, but ranges from 1.8 to 3.6 due to depolarization in the plane and being near the noise level.
\citet{2004mim..proc...63R} 
 discuss zero-level errors in WMAP, and conclude, using T-T plots, that $\beta$( 1420 MHz - 22.8 GHz) decreases from 3.1 to 2.8 after correction, nearer to the 408-1420 MHz value.

\end{appendix}

\end{document}